%% file: speed_ai.tex
\documentclass[a4paper,fleqn]{cas-sc}

\usepackage[numbers]{natbib}

\usepackage{lipsum}

\usepackage{my_settings}

\begin{document}
\let\WriteBookmarks\relax
\def\floatpagepagefraction{1}
\def\textpagefraction{.001}
\shorttitle{Street design and driving behavior: evidence from a large-scale study in Milan, Amsterdam, and Dubai}
\shortauthors{G. Orsi~al.}

\title [mode = title]{Street design and driving behavior: evidence from a large-scale study in Milan, Amsterdam, and Dubai}                
% \tnotemark[1,2]

% \tnotetext[1]{This document is the results of the research
%    project funded by the National Science Foundation.}

% \tnotetext[2]{The second title footnote which is a longer text matter
%    to fill through the whole text width and overflow into
%    another line in the footnotes area of the first page.}

\author[1,2]{Giacomo Orsi}[
   % type=editor,
                        % auid=000,
                        % bioid=1,
                        % prefix=Sir,
                        % role={Research Fellow},
                        orcid=0009-0001-8393-8090
                        ]
% \cormark[1]
\ead{gorsi@mit.edu}
% % \ead[url]{www.jkkrishnan.in}

% \credit{Conceptualization of this study, Methodology, Software}

%\address[1]{, Street 129, 1043 NX Amsterdam, The Netherlands}
\affiliation[1]{organization={MIT Senseable City Lab, Massachusetts Institute of Technology},
               %  addressline={77 Massachusetts Avenue}, 
                city={Cambridge},
%               citysep={}, % Uncomment if no comma needed between city and postcode
               %  postcode={02139}, 
                state={MA},
                country={United States}}

\affiliation[2]{organization={AMS Institute},
               %  addressline={77 Massachusetts Avenue}, 
                city={Amsterdam},
%  addressline={77 Massachusetts Avenue}, 
% city={Cambridge},
%               citysep={}, % Uncomment if no comma needed between city and postcode
%  postcode={02139}, 
% state={MA},
country={Netherlands}}

\affiliation[3]{organization={Istituto di Informatica e Telematica del CNR},
% addressline={77 Massachusetts Avenue}, 
city={Pisa},
%               citysep={}, % Uncomment if no comma needed between city and postcode
% postcode={02139}, 
% state={MA},
country={Italy}}

\affiliation[4]{organization={École Polytechnique Fédérale de Lausanne},
               %  addressline={77 Massachusetts Avenue}, 
                city={Lausanne},
%  addressline={77 Massachusetts Avenue}, 
% city={Cambridge},
%               citysep={}, % Uncomment if no comma needed between city and postcode
%  postcode={02139}, 
% state={MA},
country={Switzerland}}

\affiliation[5]{organization={Politecnico di Milano},
               %  addressline={77 Massachusetts Avenue}, 
                city={Milano},
%  addressline={77 Massachusetts Avenue}, 
% city={Cambridge},
%               citysep={}, % Uncomment if no comma needed between city and postcode
%  postcode={02139}, 
% state={MA},
country={Italy}}

% \author[2,4]{Han Thane}[style=chinese]

\author[1,2]{Titus Venverloo}[%
   % role=Co-ordinator,
   % suffix=Jr,
   ]

\author[1,2,4]{Andrea La Grotteria}[%
   % role=Co-ordinator,
   % suffix=Jr,
   ]

\author[1]{Umberto Fugiglando}[%
% role=Co-ordinator,
% suffix=Jr,
]

\author[1]{Fábio Duarte}[%
   % role=Co-ordinator,
   % suffix=Jr,
   ]

\author[1,3]{Paolo Santi}[%
   % role=Co-ordinator,
   % suffix=Jr,
   ]
% \ead{psanti@mit.edu}

\author[1,5]{Carlo Ratti}[%
   % role=Co-ordinator,
   % suffix=Jr,
   ]
% \ead[URL]{https://www.university.org}

% \credit{Data curation, Writing - Original draft preparation}

% \affiliation[2]{organization={World Scientific University},
%                 addressline={Street 29}, 
%                 postcode={1011 NX}, 
%                 postcodesep={}, 
%                 city={Amsterdam},
%                 country={The Netherlands}}

% \author[1,3]{T. Rafeeq}
% \cormark[2]
% \fnmark[1,3]
% \ead{t.rafeeq@example.in}
% \ead[URL]{www.campus.in}

% \affiliation[3]{organization={University of Intelligent Studies},
%                 addressline={Street 15}, 
%                 city={Jabaldesh},
%                 postcode={825001}, 
%                 state={Orissa}, 
%                 country={India}}

% \cortext[cor1]{Corresponding author}
% \cortext[cor2]{Principal corresponding author}
% \fntext[fn1]{This is the first author footnote, but is common to third
% \fntext[fn2]{Another author footnote, this is a very long footnote and
%   it should be a really long footnote. But this footnote is not yet
%   sufficiently long enough to make two lines of footnote text.}

% \nonumnote{This note has no numbers. In this work we demonstrate $a_b$
%   the formation Y\_1 of a new type of polariton on the interface
%   between a cuprous oxide slab and a polystyrene micro-sphere placed
\input{abstract}

% \begin{highlights}
% \item Research highlights item 1
% \item Research highlights item 2
% \item Research highlights item 3
% \end{highlights}


\maketitle

\input{content}

% \verb+\printcredits+ command is used after appendix sections to list 
% author credit taxonomy contribution roles tagged using \verb+\credit+ 
% in frontmatter.

\printcredits

\bibliographystyle{cas-model2-names}

\bibliography{bibliography}

\newpage

\footnotesize
\input{appendix}


% Author biography without author photo.
% Author biography. Author biography. Author biography.
% Author biography. Author biography. Author biography.
% Author biography. Author biography. Author biography.
% Author biography. Author biography. Author biography.
% Author biography. Author biography. Author biography.
% Author biography. Author biography. Author biography.
% Author biography. Author biography. Author biography.
% Author biography. Author biography. Author biography.
% Author biography. Author biography. Author biography.

% \bio{figs/cas-pic1}
% Author biography with author photo.
% Author biography. Author biography. Author biography.
% Author biography. Author biography. Author biography.
% Author biography. Author biography. Author biography.
% Author biography. Author biography. Author biography.
% Author biography. Author biography. Author biography.
% Author biography. Author biography. Author biography.
% Author biography. Author biography. Author biography.
% Author biography. Author biography. Author biography.
% Author biography. Author biography. Author biography.

% \bio{figs/cas-pic1}
% Author biography with author photo.
% Author biography. Author biography. Author biography.
% Author biography. Author biography. Author biography.
% Author biography. Author biography. Author biography.
% Author biography. Author biography. Author biography.

\end{document}

%% file: abstract.tex
\begin{abstract}
    In recent years, cities have increasingly reduced speed limits from 50 km/h to 30 km/h to enhance road safety, reduce noise pollution, and promote sustainable modes of transportation, such as walking and cycling. However, achieving compliance with these new limits remains a key challenge for urban planners, as adherence is crucial to achieving the intended benefits.
    
    This study investigates drivers' compliance with the 30 km/h speed limit (ca. 20 mph) in Milan and examines how street characteristics influence driving behavior. Using 51 million telemetry data points from vehicles, we analyze speed profiles across the city and employ an observational study to estimate the causal effect of speed limit reductions. Our findings indicate that the average vehicle speed in streets with a limit of 30 km/h is only 2.29 km/h lower than in those with a limit of 50 km/h with similar street layout. This suggests that the mere introduction of lower speed limits is not sufficient to reduce driving speeds effectively, highlighting the need to understand how street design can improve speed limit adherence. To comprehend this relationship, we apply computer vision-based semantic segmentation models to Google Street View images of Milan's streets. A large-scale analysis reveals that narrower streets and densely built environments are associated with lower speeds, whereas roads with greater visibility and larger sky views encourage faster driving. 

    To evaluate the influence of the local context on the urban design features that influence drivers' compliance, we apply the developed methodological framework to two additional cities: Amsterdam, which, similar to Milan, is a historic European city not originally developed for cars, and Dubai, which instead has developed in recent decades with a more car-centric design. The results of the analyses largely confirm the findings obtained in Milan, which demonstrates the broad applicability of the road design guidelines for driver speed compliance identified in this paper.
        
    Finally, we develop a machine learning model to predict driving speeds based on street characteristics. We showcase the model's predictive power by estimating the compliance with speed limits in Milan if the city were to adopt a 30 km/h speed limit city-wide. The tool provides actionable insights for urban planners, supporting the design of interventions to improve speed limit compliance, optimize traffic management strategies, and create safer urban environments.
\end{abstract}

%% file: content.tex
\input{introduction}

\input{methods}

\input{results}

\input{discussion}

\input{acknowledgements}

%% file: introduction.tex
\section{Introduction}

Urban speed limit reduction initiatives have gained significant attention in recent decades. Cities worldwide are striving to enhance road safety, reduce traffic-related injuries, lower noise and air pollution levels, and encourage more sustainable modes of transportation \citep{yannis_review_2024}. One key strategy to achieve these goals is the implementation of lower speed limits in urban areas.

In Europe, a growing trend towards 30 km/h city-wide adoption has emerged, with major cities like Zurich (progressively since 1991), Madrid (2018), London (2020), Barcelona (2020), Brussels (2021), Lyon (2022), and Amsterdam (2023) leading the way. Various studies have highlighted the benefits of reduced speed limits in urban areas. For instance, \cite{grundy_effect_2009} found that implementing 20 mph (approximately 30 km/h) speed zones in London was associated with a 41.9\% reduction in road casualties. 

Several additional studies over the following years, employing multiple methodologies, have confirmed the positive impact of speed limit reductions on road safety. A recent review by \cite{yannis_review_2024} provides a comprehensive overview of the measured effects of speed limit reductions for city-wide implementations in Europe. Besides the clear safety benefits, the review also reports positive environmental impacts, with a decrease in emissions by an average of 18\%, noise pollution levels by 2.5 dB, and fuel consumption by 7\%. However, air quality benefits still need further investigation, as other studies have shown that lower car speeds can lead to an increase in emissions due to the engine's lower efficiency at reduced speeds for combustion engines \citep{GRESSAI2021100018, ijerph16010149, INTPANIS201132}. 

Nonetheless, \cite{yannis_review_2024} report that city-wide adoption of a 30 km/h speed limit is associated with a modal shift towards more sustainable transportation modes, such as walking and cycling, and a reduction in car use, which could offset potential increases in emissions due to lower speeds.

The success of these initiatives depends on various factors, including public acceptance, enforcement, and the presence of complementary infrastructure such as sidewalks, cycle lanes, and public transport to support a modal shift. The varying success rates across different cities underscore the need for a comprehensive understanding of the local context when implementing speed limit reductions.

While a large body of research has focused on road safety, less attention has been paid to the causal mechanisms underlying the effect, particularly on the interventions that cities have to implement to ensure that drivers comply with the lower speed limits. Enforcement strategies, such as speed cameras and traffic police, can play a crucial role in this regard. However, the effectiveness of these measures can be limited by the severity of penalties and technical constraints, such as the availability of resources and the need for continuous monitoring.  Moreover, the mere presence of enforcement measures does not guarantee compliance, as drivers may adapt their behavior when they know they are being monitored \citep{mountain2004costing}. Additionally, enforcement measures can lead to negative acceptance by the public, as they are often perceived as punitive rather than preventive \citep{BLINCOE2006371}. In the context of Italy, speed cameras at 30 km/h have also been criticized and recently even banned by the Italian government, in an effort to remove the perception that cities are reducing speed limits to increase revenue from fines \citep{italian_ministry}. Cities can still enforce limits at 30 km/h with traffic police, with fines ranging from 42 € (for exceeding the limit by more than 5 and less than 11 km/h) to 173 € (for exceeding the limit by more than 11 km/h) \citep{sanzioni}. 

This highlights the need for alternative strategies to ensure drivers reduce their speed and comply with more stringent limits. Strategies include sensitization campaigns and street design interventions that encourage drivers to drive at lower speeds. 

In this study, we focus on the latter, analyzing how street design can influence driver behavior and speed compliance through a large-scale study in Milan, Italy.  To date, there is no public large-scale study that has analyzed the compliance with speed limits across entire cities, especially focusing on areas where speed limits have been reduced to 30 km/h. Most studies that focused on the relationship between street design and drivers' speed or speed limits, have been conducted in controlled or simulated environments \citep{theeuwes_self-explaining_2024, yao_close_2020}. 
While other studies have used Street View Imagery (SVI) to analyze the built environment and its association with various urban phenomena, such as the perceived value of built environment \citep{salazar_miranda_desirable_2021}, predicting travel patterns \citep{goel_estimating_2018}, real estate evaluation \citep{real_estate_eval}, understanding crime \citep{he2017built}, and mapping infrastructure defects \citep{8287715}, we introduce a use of SVI to estimate the effect of speed limit reductions on driving speeds and investigating how street design influences driver behavior and speed compliance.

%% file: methods.tex
\section{Data and Methods}
\subsection{Vehicles Speed Data}
\label{sec:data}
We use a dataset collected by UnipolTech, which operates OBU (On Board Units) on about 4.2 million vehicles in Italy \cite{UnipolTech}, collecting travel data in a non-controlled setting. Data is anonymized by UnipolTech. 
The dataset includes the speed and the heading angle of the vehicles, along with their position obtained through a GNSS receiver, collected at regular intervals (with an average sampling time of 2 minutes). The data used for this study consists of speed observations in Milan, Italy\footnote{Defined by the following bounding box: [45.532808, 45.385189, 9.277267,  9.065781]} totaling 50.98 million data points in the January, April, July, and October 2023. 

We match each street observation to the closest road segment in the OpenStreetMap (OSM) network \cite{OpenStreetMap}, accounting for the heading angle of the vehicle, as described in \ref{sec:map_matching}, and we aggregate all the speed observations on each road segment to obtain the average speed and speed quantiles per each segment. In this study we focus on the 85th percentile (\emph{speed p85}) of the speed distribution, which is often used by traffic engineers \cite{Wilmot01011999} to assess the maximum speed at which most vehicles travel, and it is often used as a reference for setting speed limits \cite{Wilmot01011999}.
\autoref{fig:speed_profiles} shows the extracted speed profiles in Milan. 
\input{fig_speed_profiles}

\subsection{OpenStreetMap Data}
\label{sec:osm}
OpenStreetMap (OSM) is an open source initiative that provides geographic data, such as street networks, points of interest, and land use information. We use the OSM data available on December 31st, 2023, 
to extract the road network and the road type (primary, secondary, tertiary, residential) of each road segment. From the 35,606 available segments within Milan, Italy, we remove 10,933 segments (30.7\%), as they are not suitable for this analysis. 
In particular, we remove 605 motorway segments, segments with a length less than 30 meters (9901, 28.29\%) and those with a length longer than 500 meters (427, 1.70\%). 
Short segments were removed to ensure a minimum length for extracting meaningful features, as Google Street View images for very short segments would capture information from surrounding segments, leading to inaccurate results. Long segments were excluded because we extract images only from the centroid of the street segments; thus, for long segments, the images would not be representative of the entire segment. From the 24,673 remaining segments, we remove those for which we could not match any speed observation from UnipolTech and those without Google Street View images. The final dataset consists of 22,818 street segments, with a total network length of 2,738 km. 
% \later{Check the numbers}. 

\subsubsection{Features extracted from OSM}
\label{sec:osm_features}
For each street segment, we extract a set of features from OpenStreetMap. The street network graph provides the topological geometry of the lines of each street segment. From that, it is trivial to extract the \feature{length} in meters. In addition, we compute the \emph{line sinuosity}, which is defined as the ratio between the length of the edge and the distance between the two endpoints.

\[
\text{sinuosity} = \frac{\text{segment length}}{\lVert(x_d, y_d) - (x_o, y_o)\rVert}
\]

where $(x_d, y_d)$ and $(x_o, y_o)$ are the coordinates of the destination and origin nodes of the segment, respectively. In addition to the street geometry, we extract features of the street segment surroundings. In particular, we are interested in the presence of points of interest, such as traffic lights, pedestrian crossings, bus stops, tram railways, sidewalks, cycleways, and parks. We extract the number of each of these features within 20 meters of the street segment geometry. The width of street segments is not among the features provided by OSM. An estimation of the width of the street is computed using a method proposed by \cite{doi:10.1177/2399808319832612} and \cite{fleischmann_2019}. 

We extract the speed limit of each street segment from OpenStreetMap, represented as the \feature{maxspeed} attribute in each OSM edge of the street network. For segments with no speed limit information, we use the default speed limit of 50 km/h, which is the standard urban speed limit in Italy when no other limit is specified.

Finally, we label each road type as \emph{primary}, \emph{secondary}, \emph{tertiary}, or \emph{residential}, based on the OSM attribute \feature{highway}, and we extract the number of lanes of each street segment from the \feature{lanes} attribute.

\subsection{Street View Imagery}
\label{sec:street_view}

For each of the 22,818 segments we obtain two images per segment from Google Street View, one heading in the direction of the street and the other in the opposite direction. Details the pictures orientation, resolution, and other parameters of the images are provided in \ref{app:google_street_view}.

We apply the current state-of-the-art semantic panoptic model for urban scenes, OneFormer \cite{jain2023oneformer}, to the street view images. Panoptic segmentation is the computer vision task of classifying every pixel of an image into a semantic category (e.g., road, sidewalk, building, bike lane) and distinguishing between every object instance (e.g., car, pedestrian, bicycle) in the image. We use a pre-trained OneFormer model on the Mapillary Vistas dataset \cite{8237796}, which is a dataset of 25,000 street-level images with pixel-level annotations in 66 categories. OneFormer achieves SOTA mIoU \cite{INTPANIS201132} of 67.4 on the dataset \cite{jain2023oneformer}.

For each street view segment, we extract the percentage of pixels in each semantic category and the number of instances of each object category. 

\subsection{Estimating Causal Effects}
\label{sec:causal_matching}

We employ a quasi-experimental design to estimate the causal effect of speed limits on vehicle speeds with a matching approach. Consider $T_i = t$ as the treatment assigned to road segment $i$ (i.e., the speed limit was reduced to 30 km/h). In the case of a binary treatment $t\in\{0, 1\}$. $Y_i^t$ denotes the potential outcome for the street $i$ under treatment $t$. The individual treatment effect is defined as $Y_i^1 - Y_i^0$ \cite{hernan2010causal}. The \emph{average treatment effect} (ATE) is defined as $\text{ATE} = \mathbb{E}[Y_i^1 - Y_i^0]$. This study uses data from 2023, where we do not observe any relevant change in the built environment or speed limits (except for the two areas analyzed in \autoref{sec:case_study}). Therefore, we cannot compute the causal effect directly. By comparing the speed of the same road segment before and after the treatment, we can only observe one of the potential outcomes, $Y_i^{T_i}$, for each unit.

Methods for estimating the ATE when the pre-treatment and post-treatment outcomes are not observed, and in the presence of unobserved confounders, include matching methods \cite{hernan2010causal}, which aim to estimate the counterfactual outcome for each unit by comparing it with other units with similar likelihood of receiving the treatment. A popular method is the propensity score matching, which estimates the probability of receiving the treatment given the observed covariates. We can assume that the probability of setting a certain speed limit is dependent on the street layout and the built environment, which we can observe from OSM and street view images. Instead of estimating the propensity score with a regression model, we directly match the treated group with the untreated (control) observation based on the similarity of the street layout and the built environment. A treated street segment $i$ ($T_i = 1$) is matched to 5 control segments $j$ ($T_j = 0$) that are most similar to the treated segment. We compute the similarity of two segments $i$ and $j$ as the percentage of overlapping pixels in the semantic categories of their street view images. Formally, let $X_{i}^{k}$ be the class assigned by the semantic segmentation model to pixel $k$ in the street view image of segment $i$, the similarity between two segments $i$ and $j$ is defined as:

\[
\text{sim}(i, j) = \frac{1}{N} \sum_{k=1}^{N} \mathbb{I}(X_{i}^{k} = X_{j}^{k})
\]

where $N$ is the total number of pixels in the street view image, and $\mathbb{I}(\cdot)$ is the indicator function. $sim(i, j)$ is a value between 0 and 1, where 0 indicates no similarity between the two segments, and 1 indicates that the two segments are identical, in their semantic classification.

Additionally, we filter out the control segments with a road type (as defined in OpenStreetMap: primary, secondary, tertiary, residential) different from the treated segment, and we match each treated segment with the 5 most similar control segments. 

We then estimate the ATE as the mean difference in the average speed of the treated and control segments. 

\input{fig_causal_matching.tex}

%% file: fig_speed_profiles.tex
\begin{figure}[ht]
    \centering

    \begin{minipage}
        {.49\linewidth}
        \centering
        \begin{overpic}[width=\linewidth,abs,unit=1mm]{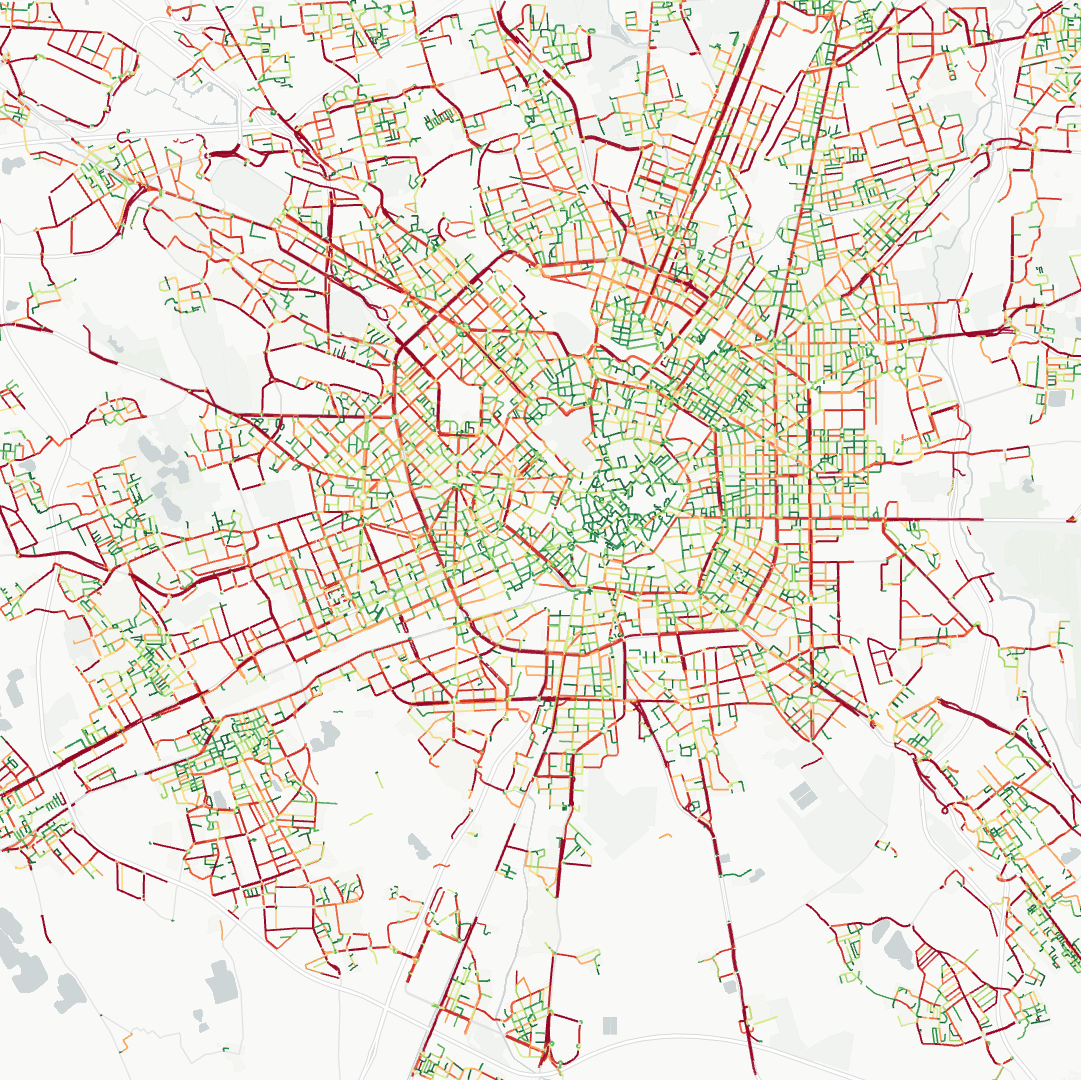}
            \put(.01\linewidth,.55\linewidth){\includegraphics[width=16mm]{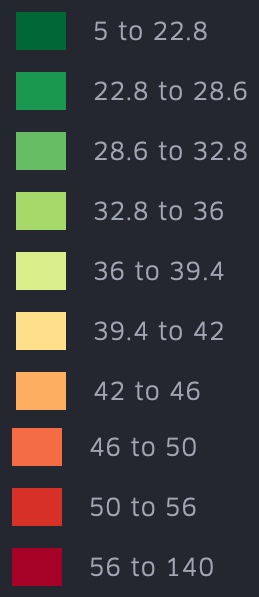}}
            \put(0.8\linewidth,.01\linewidth){\includegraphics[width=13mm]{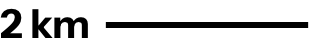}}
        \end{overpic}
    \end{minipage}
    \caption{85th percentile speed profiles of the City of Milan.}
    \label{fig:speed_profiles}
\end{figure}

%% file: fig_causal_matching.tex
\begin{figure*}[h]
    \centering
    \includegraphics[width=\linewidth]{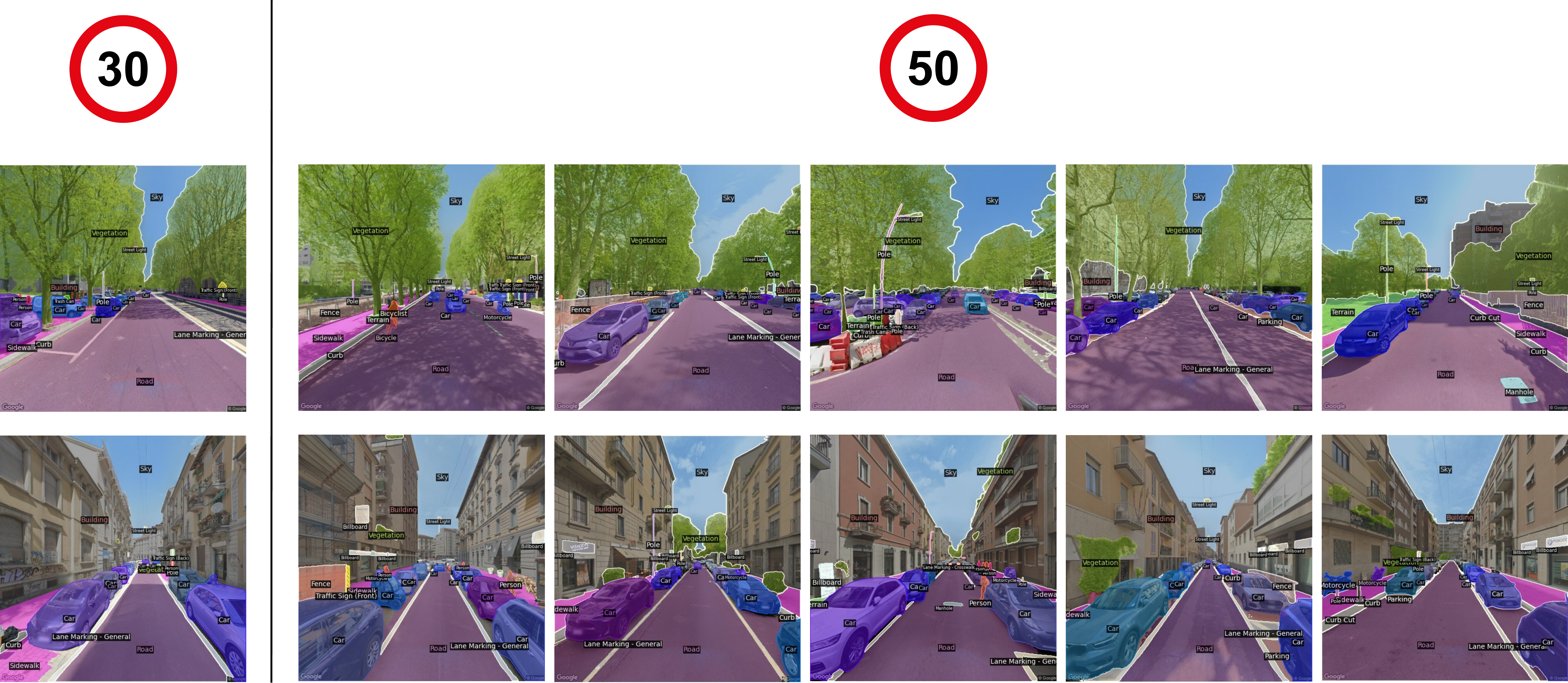}
    \caption{To estimate the causal impact of speed limits reduction, we matched every 30 km/h segment with the five most similar 50 km/h segments based on the street layout and built environment. The figure shows an example of two 30 km/h segments (left) and their five matched 50 km/h segments (right).}
    \label{fig:causal_matching} 
\end{figure*}

%% file: results.tex
\section{Results}
\subsection{Compliance with Speed Limits}
\autoref{fig:speed_profiles} shows the observed speed 85th percentile in Milan. 

By focusing on the streets with a limit at 30 km/h, we identify that the average speed 85th percentile in these areas is 38.26 km/h, with an average speed of 26.98 km/h. This indicates an overall trend of non-compliance with the speed limit, which we depict in \autoref{fig:zones_30_hour_distribution}, where we analyze the distribution per hour of the day. The data indicates a clear pattern of elevated speeds during the night, peaking between 2:00 and 3:00, when the traffic volume also decreases. Conversely, the lowest speeds are recorded during the morning (8:00-9:00) and afternoon (15:00-17:00) periods, when streets are busier.
\input{fig_zones_30_hourly}

\subsubsection{Spatial Distribution of Speed}
We quantify the spatial autocorrelation of vehicle speeds using Moran’s Local  $I$ , a statistical measure that evaluates the degree of clustering or spatial dependency of similar speed values within the urban landscape \citep{LISA}. This method helps to identify the presence of spatial patterns and relationships in the data. 

For each 30 km/h segment, we select the $K$ road segments within a 300-meter radius and compute the Moran's local $I$ using the formula in \autoref{eq:moran_local_i} for each segment.

\begin{equation}
\label{eq:moran_local_i}
I_{i}={\frac {x_{i}-{\bar {x}}}{\sigma^2}}\sum _{j=1}^{K}w_{ij}(x_{j}-{\bar {x}})
\end{equation}

where \(\sigma^2 = \frac{\sum_{i=1}^N (x_i - \bar{x})^2}{N}\) is the observed variance of the data, \(x_i\) is the speed 85th percentile of road segment \(i\), \(\bar{x}\) is the mean speed p85, \(x_j\) for \(j=1, \dots, K\) is the speed p85 of the \(j\)-th neighbor of road segment \(i\), \(w_{ij}\) is the normalized length of the \(j\)-th segment (\(\forall i \, \sum_{j=1}^K w_{ij} = 1\)), and \(N\) is the total number of road segments.

The formula measures the degree of similarity between the speed p85 of the target road segment and its neighbors, adjusted by the mean and variance of the data. A positive Moran's local $I$ indicates a cluster of similar values around a target road segment, while negative values suggest a dispersed pattern.

We also compute a Global Moran's $I$, which measures the overall spatial autocorrelation of the data. This is obtained by averaging the local Moran's $I$ values for all road segments, weighted by the length of each road segment, as detailed in \autoref{eq:moran_global_i}, where \(l_i\) is the length of road segment \(i\):

\begin{equation}
\label{eq:moran_global_i}
I = \frac{1}{\sum_{i=1}^N l_{i}} \sum_{i=1}^N l_{i} I_{i}
\end{equation}

\begin{figure}[h]
    \centering
    \begin{subfigure}[b]{0.49\linewidth}
        \centering
        \includegraphics[width=\linewidth]{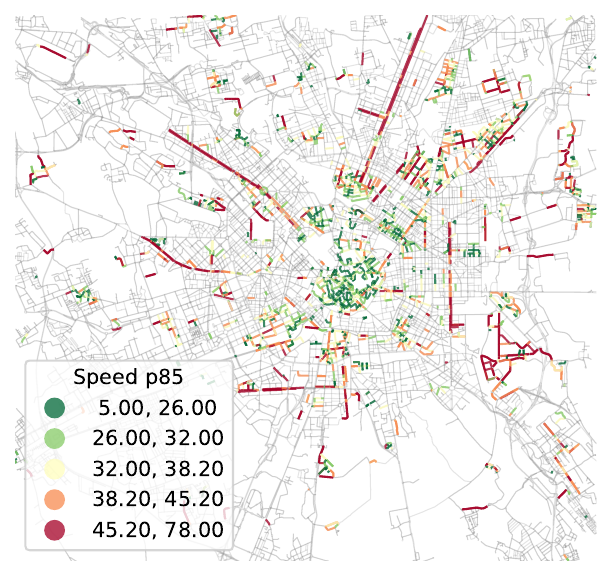}
        \subcaption{speed p85 in 30 km/h zones}
    \end{subfigure}
    \hfill
    \begin{subfigure}[b]{0.49\linewidth}
        \centering
        \includegraphics[width=\linewidth]{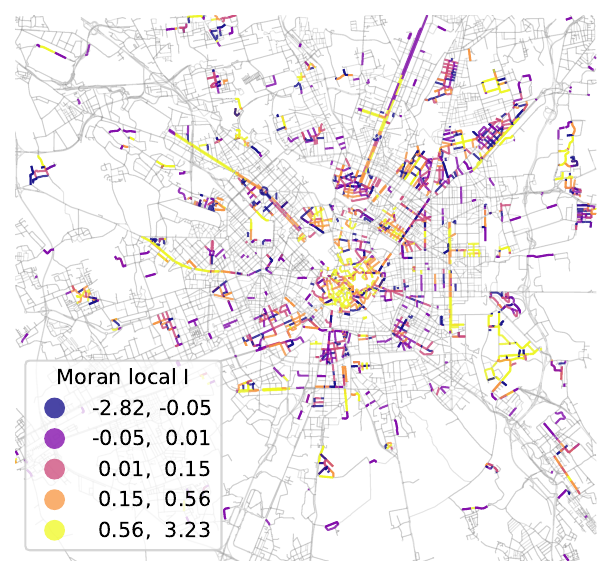}
        \subcaption{Local Moran's I in 30 km/h zones}
    \end{subfigure}
    \caption{Spatial distribution of speed p85 Zones 30}
    \label{fig:zones_30_spatial_autocorrelation}
\end{figure}

This results in a score of 0.38, indicating positive but limited overall spatial autocorrelation for speed limit compliance in the 30 km/h zones. \autoref{fig:zones_30_spatial_autocorrelation} illustrates the spatial distribution of the speed p85s and the local Moran's $I$. The map reveals clusters of roads with similar drivers' speed,  with a notable concentration of high compliance in the historical center of Milan. In other areas, the map displays mixed patterns of spatial correlation, where high-compliance streets are surrounded by low-compliance ones. This suggests that the spatial distribution of speed in 30 km/h zones may be influenced by localized factors such as road design, traffic volume, POIs, and urban density, rather than being uniformly distributed in each neighborhood. 

\subsection{Causal Impact of Speed Limits Reduction}
\label{sec:results_speed_limits}

In this section we estimate the causal impact of speed limits reductions from 50 km/h to 30 km/h in two ways: first with a before-after analysis in two areas of Milan, and second with a city-wide observational study where we match 30 km/h segments with 50 km/h segments with similar street layout and built environment. 

\subsubsection{Before-and-after case study in two areas}
\label{sec:case_study}
In 2023, the City of Milan reduced the speed limit from 50 km/h to 30 km/h in two urban areas\cite{milano_dataset}: the neighborhoods of Porta Volta and Isola (see \autoref{fig:case_study_areas}). For each area, we collected two datasets of speed observations from UnipolTech: one covering three weeks starting one month before the speed limit reduction and another covering three weeks starting one week after the speed limit reduction. 
In Porta Volta, the average vehicle speed was 23.33 km/h before the speed limit reduction and 22.70 km/h after. In Isola, the average speed was 25.48 km/h before the speed limit reduction and 25.70 km/h after. We compare the speed distributions before and after with a $t$-test, which for Porta Volta resulted in a $p$-value of 0.11, and for Isola, a $p$-value of 0.051. The percentage of vehicles exceeding 30 km/h Porta Volta was 12.19\% before the speed limit reduction and 13.29\% after. In Isola, the corresponding values were 13.02\% and 13.29\%. 
The interesting results indicate that vehicle speeds did not significantly change in either area following the speed limit reduction, as the slight variations in average speed are not statistically significant. 
\input{fig_case_study_areas}

\subsubsection{City-wide Observational Study of Speed Limits Reduction Effect}
\label{sec:observational_study}

As we could observe before-and-after speed only in two areas of Milan, we conduct a city-wide observational study to estimate the causal impact of speed limits reduction from 50 to 30 km/h on vehicles speed in all other areas, by matching 30 km/h segments with 50 km/h segments with similar street layout based on the segmentation masks of Google Street View images, using the approach described in \ref{sec:causal_matching}. 

We matched each of the 1694 segments with a limit at 30 km/h to 5 segments with a limit at 50 km/h, as described in Section \ref{sec:causal_matching} and shown in \autoref{fig:causal_matching}. The average vehicle speed in the 30 km/h segments was 27.19 km/h, and in the matched 50 km/h segments was 29.48 km/h. This results in an estimated causal effect of -2.29 km/h (two-sample $t$-test $p$-value < 0.001). When considering the 85th percentile of the speed distribution, the average speed 85th percentile in the 30 km/h segments was 38.73 km/h, while in the 50 km/h segments was 41.97 km/h. This results in an estimated causal effect of -3.45 km/h (two-sample $t$-test $p$-value < 0.001).

\subsubsection{Findings in Causality}
\label{sec:causal_discussion}

The results of the before-and-after study and the observational study suggest that the speed limit reduction from 50 km/h to 30 km/h in the City center of Milan does not have a significant impact on vehicle speeds. There is not statistically significant evidence that the speed limit reduction led to a decrease in vehicle speeds in the two areas analyzed in the before-and-after study – where the city reduced the limits in 2023. Additionally, the results shown by the observational study suggest that the speed limit reduction has a small effect on vehicle speeds, with an estimated $-3.45$ km/h in the 85th percentile of the speed distribution. 

These findings suggest that the mere introduction of lower speed limits is not sufficient to reduce driving speeds effectively, highlighting the need to understand how street design, or other strategies such as enforcement and sensitization campaigns, can improve speed limit adherence. The results justify the analysis carried out in the following sections, where we investigate how street design influences driver behavior and speed compliance.

Limitations of this approach could be due to the presence of other confounding factors, which might both influence the decision to reduce the speed limit and the speed at which vehicles travel. Examples of such factors that might not be captured by the matching method proposed include the vicinity of schools, hospitals, or pedestrian areas, and traffic. The before-and-after study results, and the relatively small estimated causal effect, suggest that the City might have reduced the speed limits in areas where vehicles were already traveling at lower speeds, indicating that it is not necessarily the speed limit that influences vehicle speeds, but other factors such as street design. 

\section{Association of street features and speed}
\label{sec:correlation_analysis}

To inform urban planning strategies that promote speed limit compliance at 30 km/h zones, we analyze the relationship between street design features and vehicle speeds. We investigate how physical road characteristics and visual environmental cues influence driver behavior and speed limit adherence.

We divide the street segments of the currently existing streets with a limit at 30 km/h into two equally-sized group based on the median of the observed speed 85th percentile, corresponding to 39.43 km/h. The average speed for the high-compliance group is 18.53 km/h, while the low-compliance group has an average speed of 31.54 km/h. The spatial distribution of the two groups is shown in Figure \ref{fig:zones_30_compliance_map}. 
\input{fig_compliance_map}

For each relevant feature \(f\), we report the difference in means \(d_f\) between the two groups, as detailed in \autoref{eq:mean_difference}. We also employ a Mann-Whitney U test to test the null hypothesis that the two groups have the same distribution of the feature and report the features with the smallest $p$-values in \autoref{fig:zones_30_osm_features} and \autoref{fig:zones_30_gsv_features}.
\input{fig_zones_30_features_comparison}

\begin{equation}
\label{eq:mean_difference}
    d_f = \mu^f_{\text{\ high compliance}} - \mu^f_{\text{\ low compliance}}
\end{equation}

The width of the road, measured by \feature{building\_width}, indicates that narrower streets are associated with higher compliance ($d = -12.81$ meters). This is further supported by features extracted from Google Street View (GSV) images, which relate to the driver's view of the road. Greater road visibility (measured by the percentage of road pixels in the image) correlates to lower compliance, suggesting that roads perceived as open and unobstructed encourage faster driving ($d = -0.04\%$ pixels). Conversely, higher proportions of building pixels ($d = 0.10\%$ pixels) are associated with denser urban environments and better compliance, reinforcing the impact of urban density on driver behavior.

Longer road segments ($d = -49.23$ meters), which can also be interpreted as the distance between intersections, appear to encourage higher speeds. The number of lanes ($d = -0.19$ lanes) is another significant factor, with fewer lanes correlating with higher compliance. This suggests that reducing speed limits on roads with multiple lanes might be ineffective in reducing speeds, as drivers may not comply with the new limits.

One-way streets show a complex relationship with compliance ($d = 0.20$ on binary variable): while they are generally associated with higher compliance, there is also a significant number of one-way streets with low compliance. This suggests that the effect of one-way streets on speed compliance is not uniform and may depend on other factors. Therefore, changing the speed limit on all one-way streets indiscriminately, as some policies (e.g., by the Spanish government \citep{etsc_spain}) propose, may not be effective in Milan. Urban planners should consider other road layout factors as well.

Interestingly, the presence of bus stops and public transport platforms ($d = 0.50$ on binary variable) is negatively correlated with speed limit adherence. An explanation for this counterintuitive finding may be that public transport facilities are often located on busier, wider, and longer arterial roads, which may be designed for higher speeds.

In conclusion, integrating physical road characteristics with visual environmental features is significantly associated with speed limit compliance in Zones 30. Urban planners and traffic managers can use these insights to design safer, more compliant urban environments. Strategies such as enhancing urban density, introducing traffic calming measures, and incorporating visual cues can collectively improve adherence to speed limits, ultimately enhancing road safety and urban livability. Future research should continue exploring these relationships.

\section{Predicting speed limit compliance based on street features}
\label{sec:machine_learning}

Developing a machine learning model to predict speed limit compliance using machine learning techniques is both scientifically and practically significant. Scientifically, it deepens our understanding of the interplay between urban features and driver behavior. Practically, it aids urban planning, especially in the effective implementation of city-wide 30 km/h zones.  By identifying areas where reduced speed limits might not achieve desired compliance, urban planners can make more informed decisions, potentially avoiding ineffective interventions and concentrating resources where they are most needed.
 
We predict a speed limit compliance score, defined as the ratio of the speed 85th percentile to the speed limit, using OSM features and the percentage of pixels in each segmentation class extracted from GSV images, using machine learning models. Additionally, we include the approximate traffic density, a feature that proxies for congestion, road capacity, and road importance, defined as the number of speed observations in each segment, divided by the segment length. The speed limit compliance score is equal to or lower than 1 when 85\% of the vehicles are driving below or at the speed limit, and higher than 1 when 85\% of the vehicles are driving above the speed limit.

We report the results with the best performance on the test set, the Gradient Boosting Regressor (details on other models are available in \autoref{app:results}). 

When the approximate traffic density is included, the model achieves an \(R^2\) score of 0.719 and a Mean Absolute Error (MAE) of 0.119, indicating strong predicting power. Excluding this feature results in a lower \(R^2\) score of 0.621 and an MAE of 0.139. 

Unexplained variance in the model's predictions could come from factors not included in the features, such as temporal variations (e.g., time of day, weather conditions), individual driver characteristics, unmeasured environmental influences, and traffic congestion, which significantly affects speed behavior. Despite these limitations, the model's performance is strong enough to offer valuable insights and predictions.

\subsection{Inference on current zones 50}
Like many other European cities, the City of Milan is considering extending the 30 km/h policy to additional areas \citep{milano_citta_30}. To assess the utility of our model, we train it on data from existing 30 km/h zones and apply it to streets currently limited at 50 km/h. The goal is to identify which of these streets drivers would likely comply with the reduced speed limit and which ones would require additional interventions, such as design modifications, traffic calming measures, or enforcement.

The inference dataset comprises 16,456 road segments with a 50 km/h limit. From the total of 19,403 road segments with a limit at 50 km/h, we exclude streets classified by OSM as primary or secondary roads, where reducing the speed limit to 30 km/h might not be advisable \citep{swiss_study}. The model, trained on the current 30 km/h zones, achieves an \(R^2\) value of 0.646 on the validation set, with a MAE of 0.173, corresponding to a 5.19 km/h error in the 85th percentile speed. This lower performance compared to the model previously discussed is likely due to a smaller training set and a reduced variance in the target variable, making prediction more challenging.

\autoref{fig:inference_map} shows the predicted speed limit compliance score for 50 km/h streets. Urban planners can use this map to identify which areas might successfully transition to a 30 km/h limit and which might require additional measures. 
\input{fig_inference_map}

For the 1,959 km of streets in the inference dataset, the model predicts the following:
\begin{itemize}
    \item{335 km (3,743 street segments, 17.09\% of the total length) are predicted to have speed 85th percentile lower than 30 km/h. For these streets, reducing the speed limit to 30 km/h would likely be successful without additional interventions or extra costs for the municipality.}
    \item{796 km (7,535 street segments, 40.64\% of the total length) are predicted to have speed 85th percentile between 30 km/h and 40 km/h. For these streets, a reduction to 30 km/h should be accompanied by additional measures to ensure compliance.}
    \item{828 km (5,178 street segments, 42.27\% of the total length) are predicted to have an 85th percentile above 40 km/h. For these streets, a reduction to 30 km/h is likely to be unsuccessful without additional interventions. The municipality should either consider significant changes to these streets, implement enforcement strategies, or exclude them from the speed limit reduction policy.}
\end{itemize}

\autoref{fig:inference_high} and \autoref{fig:inference_low} display images of eight randomly sampled streets from the top and bottom 10\% of predicted speed p85s, respectively. These images highlight the differences in street design and infrastructure between the two groups. Streets in the high compliance category tend to have narrower roads, fewer lanes, and more visual cues encouraging slower driving, consistent with findings discussed in \autoref{sec:correlation_analysis}.

\input{fig_inference_images}

\newpage

\section{Comparison with Amsterdam}

For Amsterdam, we used data from TomTom Traffic to obtain speed distributions for each road segment, considering the same period of 2023. Similarly to Milan, speed limit compliance is obtained by considering the 85\% percentile of the speed distribution. 

\begin{figure}[h]
    \centering
    \includegraphics[width=0.7\linewidth]{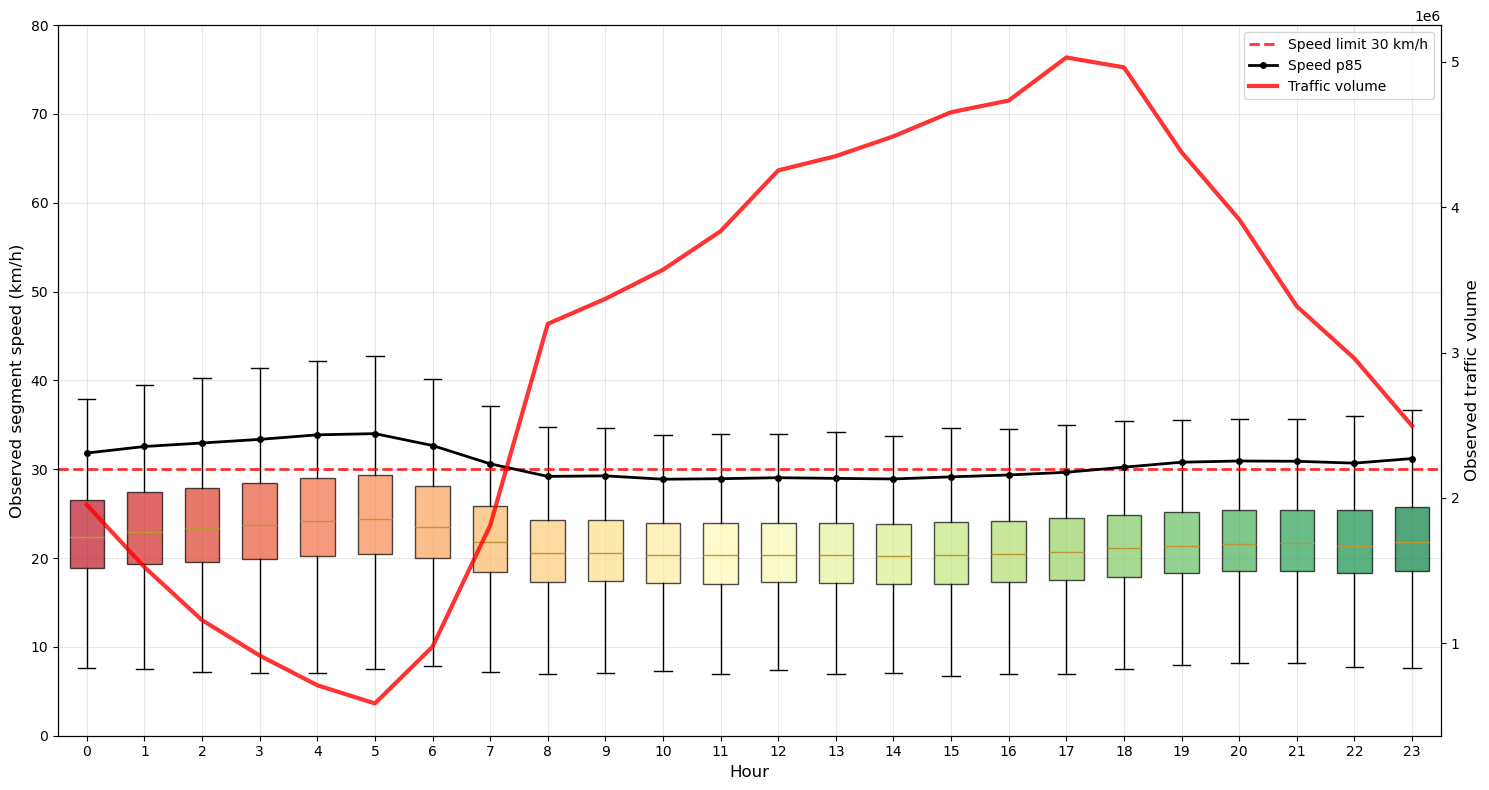}
    \caption{Speed distribution for each hour in 30 km/h streets in Amsterdam.}
    \label{fig:ams_speed30} 
\end{figure}

\begin{figure}[h]
    \centering
    \includegraphics[width=.45\linewidth]
    {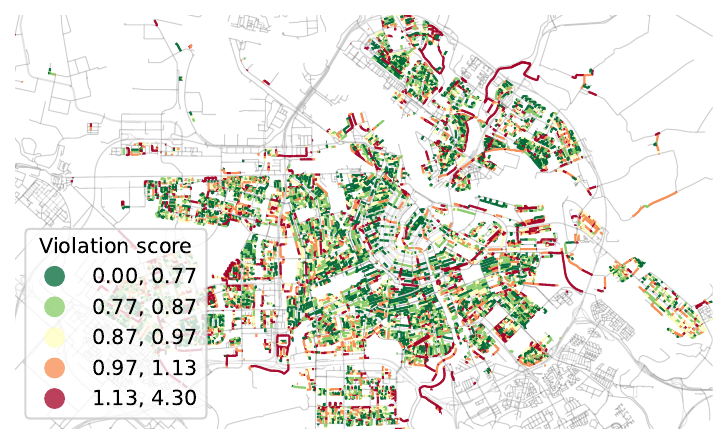}
    \includegraphics[width=.45\linewidth]{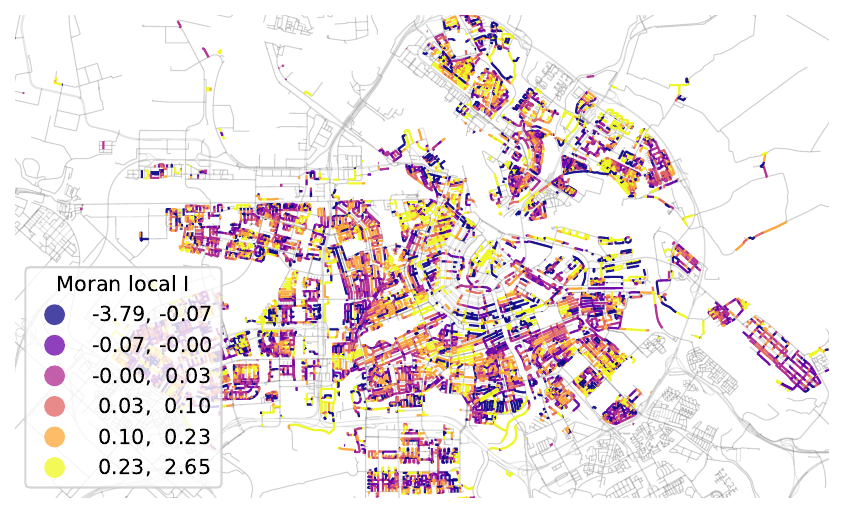}
    \caption{Spatial distribution of violation score in Amsterdam (left) and local Moran's I of violation score (right)}
    \label{fig:ams_spatial_speed} 
\end{figure}

Figure \ref{fig:ams_speed30} reports the hourly speed distribution for streets with a 30 km/h speed limit in Amsterdam. As seen from the figure, compliance with the speed limit is mostly fulfilled, except during night hours. The spatial pattern of speed limit compliance is reported in Figure \ref{fig:ams_spatial_speed}, which shows areas of non-compliance along main arteries. Note that the cluster of non-compliant streets on the north east side of the city are located in the port area, where a very low speed limit set for port operations might be the reason for the observed non-compliance. Local Moran's I computation reveals clusters of spatial autocorrelation mostly focused in the center, although noticeable clusters are observed also in the peripheral areas.

To identify road design features that influence speed limit compliance, we have used street view images obtained from Amsterdam Panorama, an open data initiative of the city of Amsterdam that regularly collects 360 views for all road segments in Amsterdam \cite{AmsPanorama}. For consistency, we used the same features from OSM and from the segmentation of Street View images used in the city of Milan.

\begin{figure}[h]
    \centering
    \includegraphics[width=\linewidth]{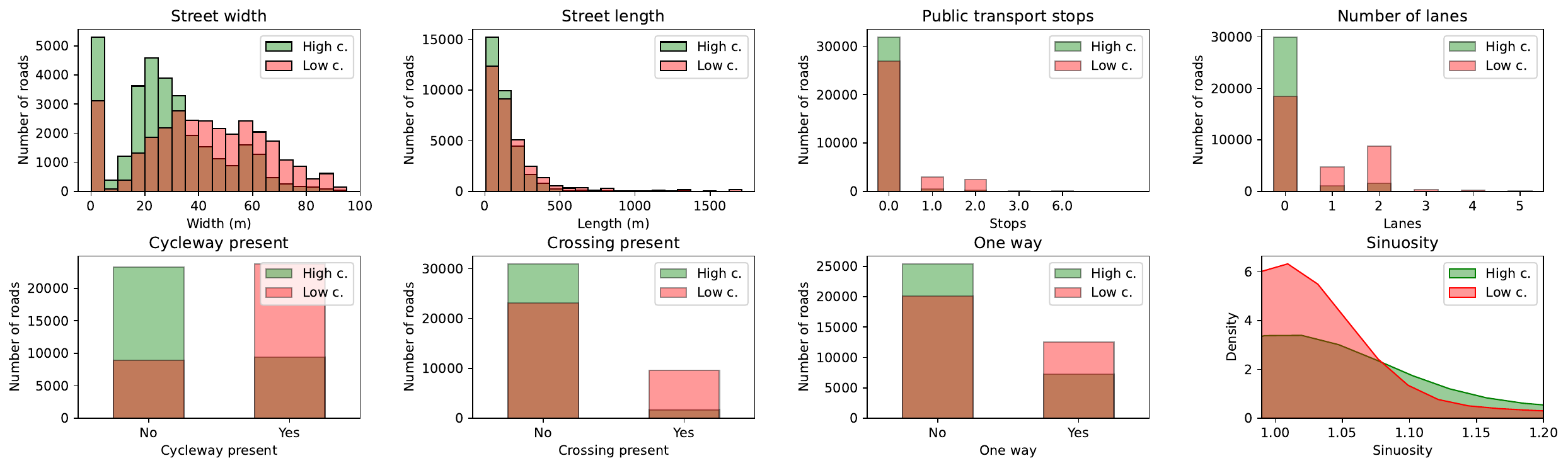}
    \caption{OSM features comparison between high and low compliance streets in Amsterdam}
    \label{fig:ams_features} 
\end{figure}

The results of the analysis are reported in Figure \ref{fig:ams_features}, and largely confirm the effects of road features on speed limit compliance observed in Milan: streets that are narrower, shorter, have fewer public transit stops and number of lanes tend to reduce driving speed, hence improve speed limit compliance. The effect of road sinuosity is also consistent with what is observed in Milan. Other features are specifically observed in Amsterdam, such as a lack of visible bike lanes, as most bike lanes are separated from car lanes, which tends to reduce speed. This might be due to a cross-correlation between the presence of bike lanes and overall street width, and the observed effect of street width on speed.

%Finally, we used data from the Municipality on the streets that underwent an street layout intervention.

% \begin{figure}[h]
%     \centering
%     \begin{subfigure}[b]{0.45\linewidth}
%         \centering
%         \includegraphics[width=0.93\textwidth]{amsterdam_img/zones_30_avg_speed.pdf}
%         \subcaption{\footnotesize Speed distribution of Zones 30 with high and low compliance in Amsterdam}
%         \label{fig:compliance_speed_comparison_ams}
%     \end{subfigure}
%     \begin{subfigure}[b]{0.3\linewidth}
%             \centering
%             \includegraphics[width=0.95\textwidth]{amsterdam_img/zones_30_map_compliance.pdf}
%             \subcaption{\footnotesize Map of the areas with high and low compliance in Zones 30.}
%             \label{fig:zones_30_compliance_map_ams}
%     \end{subfigure}
%     \caption{Speed distribution and compliance map in Zone 30 in Amsterdam}
%     \label{fig:compliance_map_ams}
% \end{figure}

% \begin{figure}[h]
%     \centering
%     \includegraphics[width=.45\linewidth]
%     {amsterdam_img/zones_30_avg_speed.pdf}
%     \includegraphics[width=.45\linewidth]{amsterdam_img/zones_30_map_compliance.pdf}
%     \caption{Spatial distribution of speed p85 in Amsterdam (left) and local Moran's I of speed p85 (right)}
%     \label{fig:ams_spatial_speed} 
% \end{figure}

\begin{figure}[h]
    \centering
    \includegraphics[width=\linewidth]{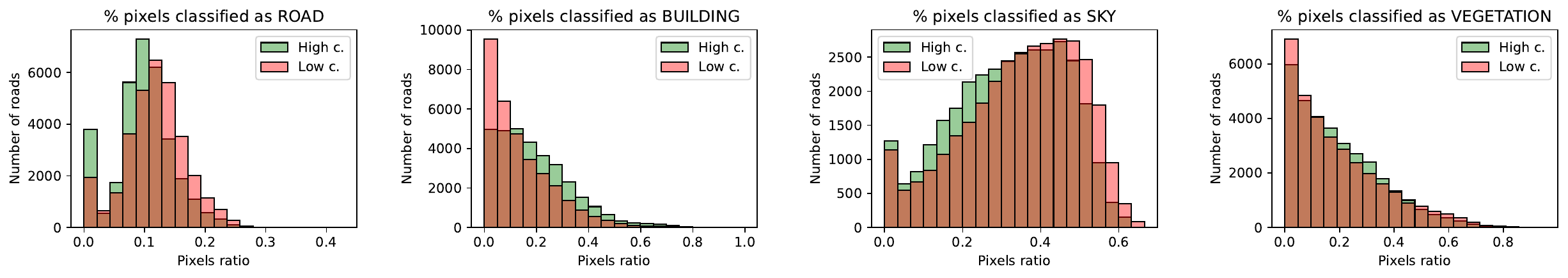}
    \mycaption{GSV features comparison between high and low compliance Zones 30 in Amsterdam}{We report the features with the lowest $p$-values in the Mann-Whitney U test. All $p$-values are below 0.001}
    \label{fig:ams_features_gsv} 
\end{figure}

% \includegraphics[width=\linewidth]{amsterdam_img/features_importance_xgboost.png}

%\includegraphics[width=\linewidth]{amsterdam_img/speed_prediction.png}

%\includegraphics[width=\linewidth]{amsterdam_img/before_after_limit_change.png}

%\includegraphics[width=\linewidth]{amsterdam_img/speed_change_intervention.png}

% When the approximate traffic density (\autoref{sec:approximate_traffic_volume}) is included, the model achieves an \(R^2\) score of 0.697 and a Mean Absolute Error (MAE) of 0.127, indicating strong predicting power. Excluding this feature results in a lower \(R^2\) score of 0.593 and an MAE of 0.145. This variable is a proxy for congestion, road capacity, and road importance: those are factors that significantly influence speeding behavior. We focus the rest of the analysis on the model that includes the approximate traffic density (more results on both models are available in \autoref{app:results}).

% We employ different machine learning models to predict street segments average speed based on the street design features. We compare the performance of linear regression, random forest, and gradient boosting models in \autoref{tab:ml_results}. The gradient boosting model outperforms the other models, achieving an $R^2$ score of 0.72. The high predictive power of the model proves the strong association between street design features and vehicle speeds, as discussed in \autoref{sec:correlation_analysis}.

\newpage

\section{Comparison with Dubai}

For Dubai, we used TomTom Traffic data to obtain speed distributions for each road segment. Due to data availability, we focused on three zones: Downtown, DIFC, and Jumeirah.
As in Milan and Amsterdam, speed limit compliance was assessed using the 85th percentile of the speed distribution.

\begin{figure}[h]
    \centering
    \includegraphics[width=0.6\linewidth]{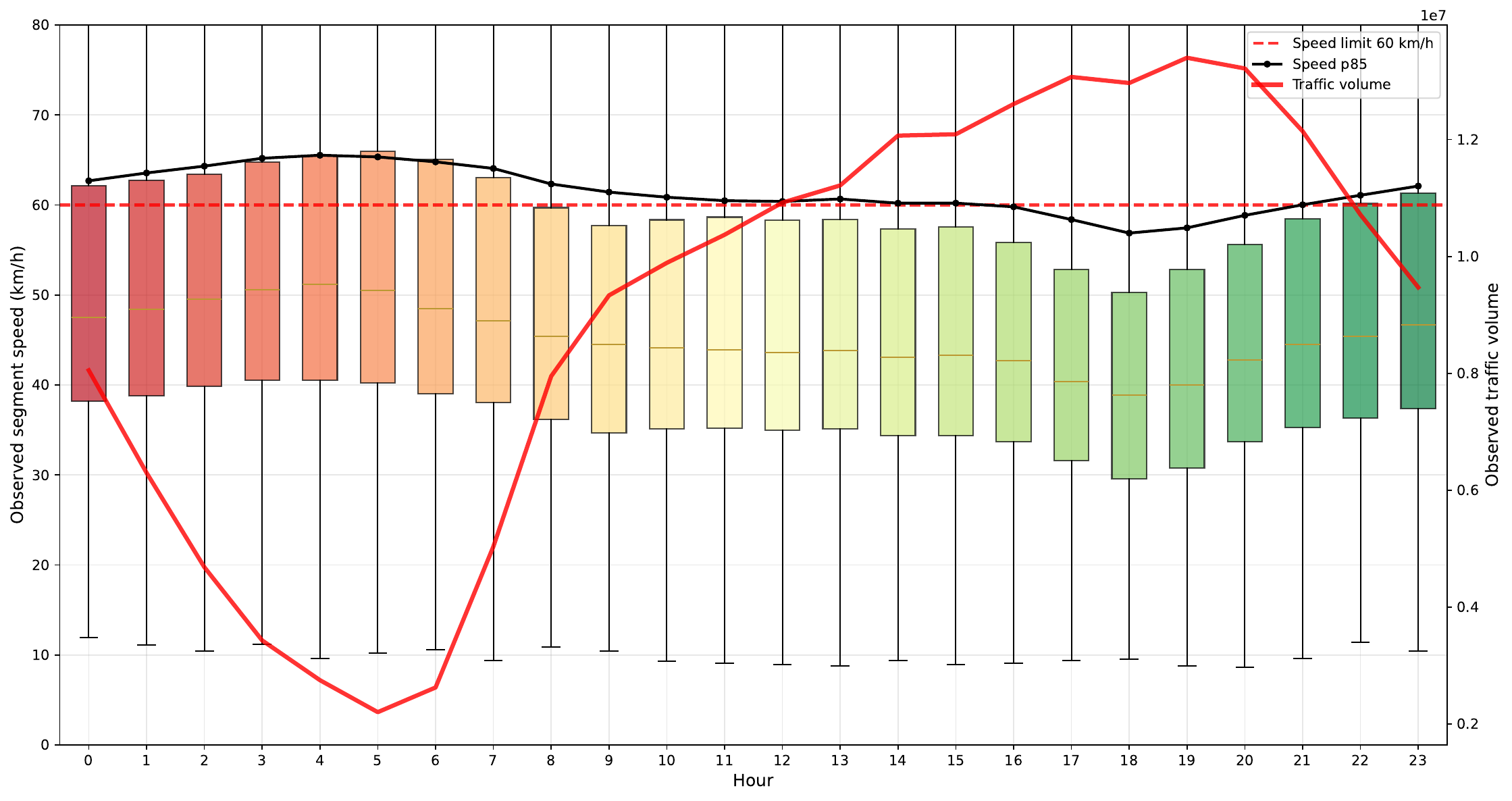}
    \caption{Speed distribution for each hour in 60 km/h streets in Dubai.}
    \label{fig:dubai_speed60} 
\end{figure}

\begin{figure}[h]
    \centering
    \includegraphics[width=.4\linewidth]
    {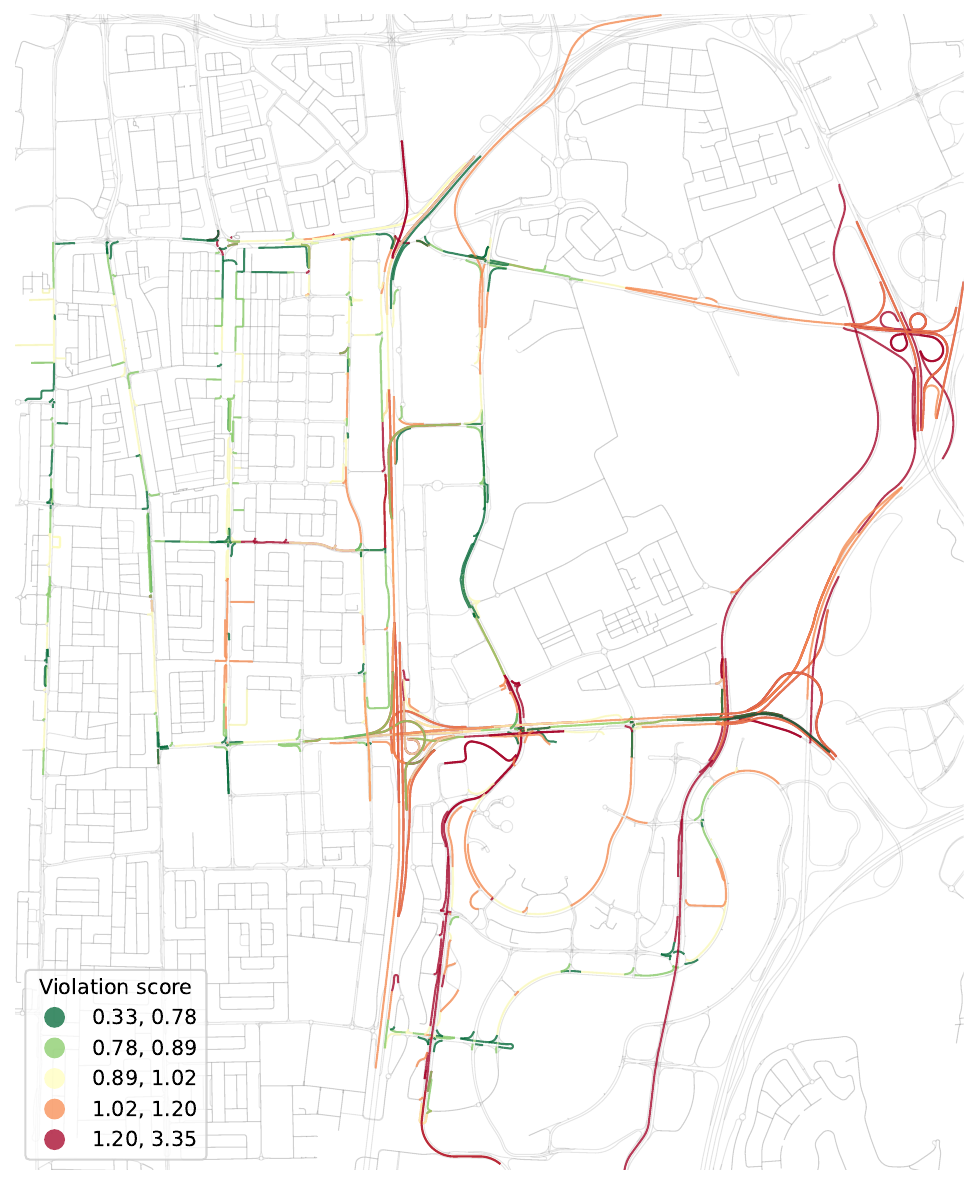}
    \includegraphics[width=.4\linewidth]{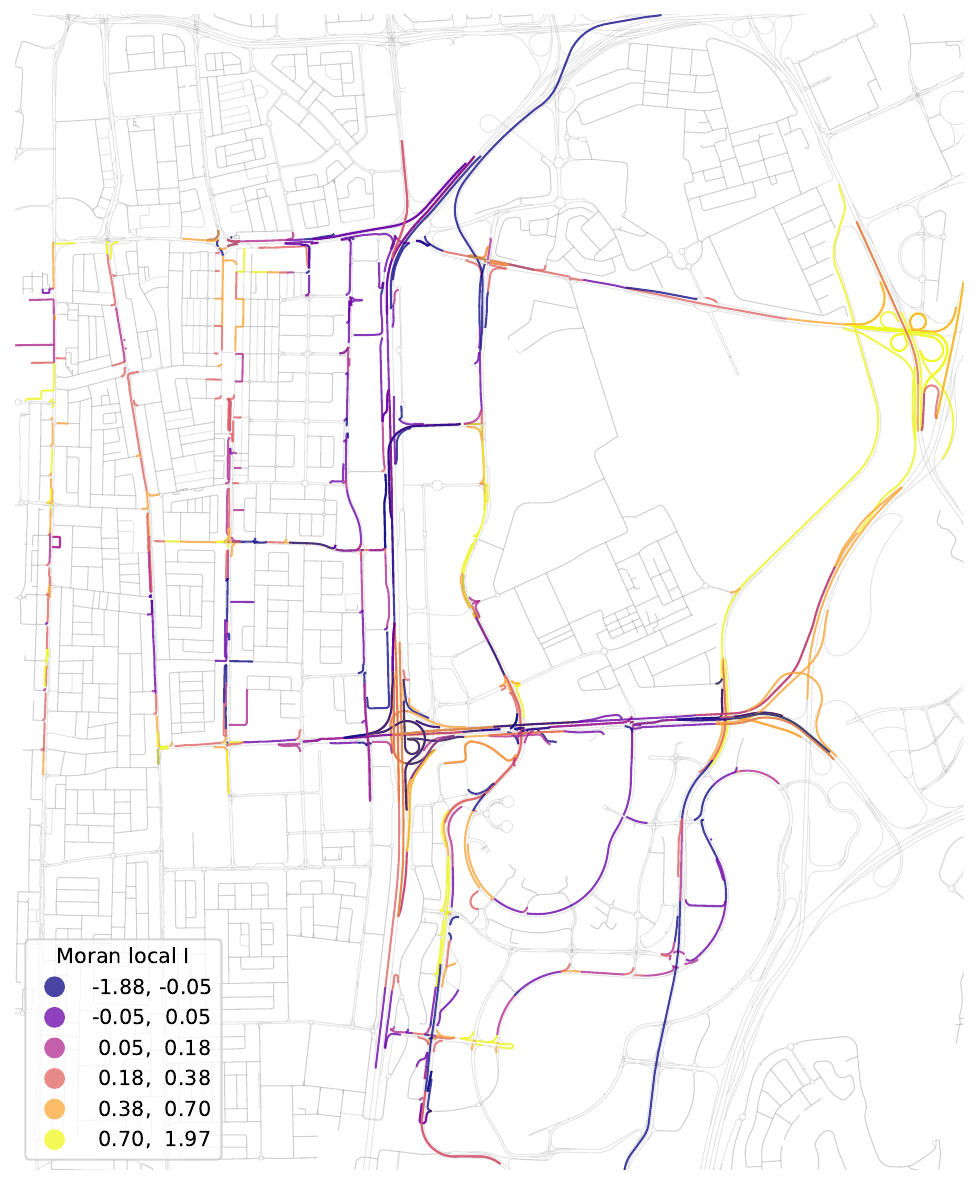}
    \caption{Spatial distribution of violation score in Dubai (left) and local Moran's I of violation score (right)}
    \label{fig:dubai_spatial_speed} 
\end{figure}

\begin{figure}[h]
    \centering
    \includegraphics[width=\linewidth]{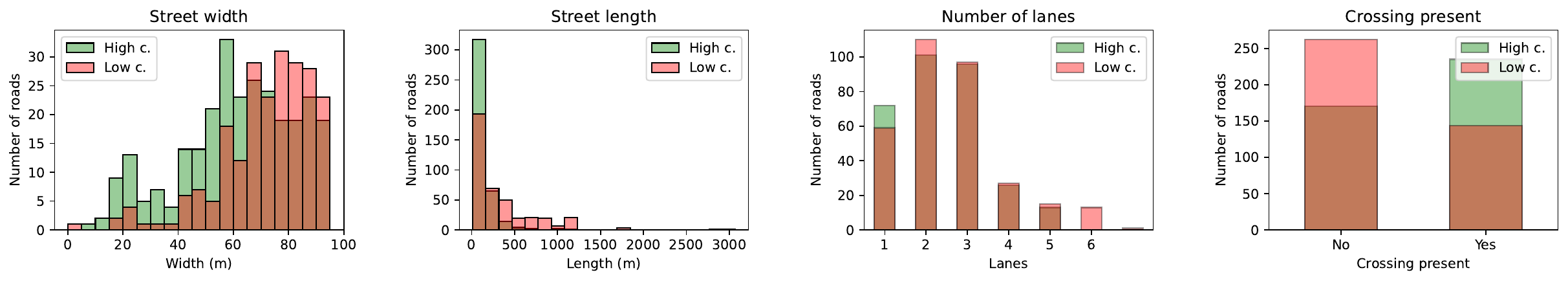}
    \caption{OSM features comparison between high and low compliance streets in Dubai}
    \label{fig:dubai_features} 
\end{figure}

To identify road design features that influence compliance, we used Google Street View images. For consistency, we relied on the same features from OpenStreetMap (OSM) and the segmentation of Street View images as in the Milan and Amsterdam analyses. Figure \ref{fig:dubai_speed60} shows the hourly speed distribution for roads with a 60 km/h speed limit in Dubai. This limit was chosen due to the limited number of roads with lower posted limits in the dataset. As observed in Milan, speeds tend to increase during nighttime hours. 
Figure \ref{fig:dubai_spatial_speed} displays the spatial distribution of compliance, highlighting clusters of non-compliance along major arterial roads. Preliminary analysis of OSM-derived features (Figure \ref{fig:dubai_features}) confirms trends found in Milan and Amsterdam: streets that are narrower, shorter, and have fewer lanes are associated with lower driving speeds. The presence of crossings within the segment also contributes to reduced speeds.

The analysis of road design using Google Street View images for the entire city of Dubai will be included in the final publication.

\newpage

%% file: fig_zones_30_hourly.tex
\begin{figure}[h!]
    \centering
    \includegraphics[width=0.8\linewidth]{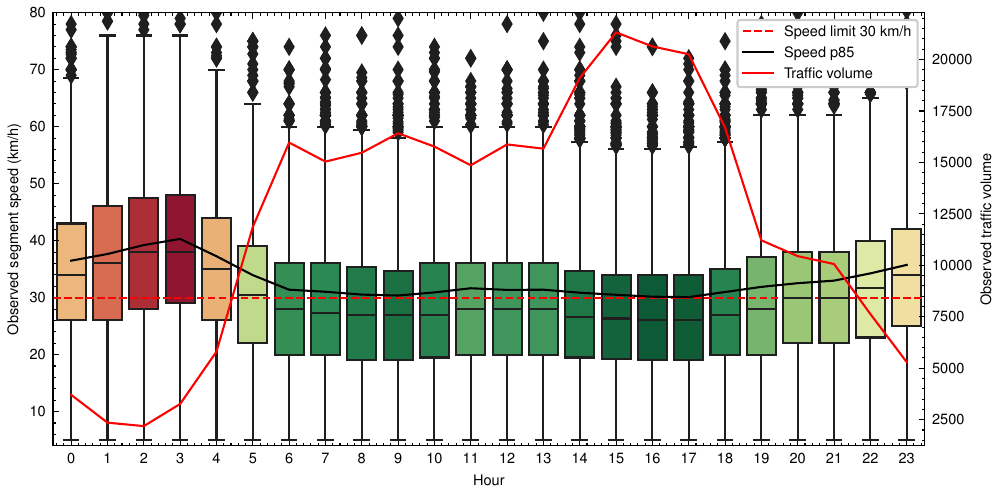}
    \caption{Speed distribution for each hour in 30 km/h streets.}
    \label{fig:zones_30_hour_distribution}
\end{figure}

%% file: fig_case_study_areas.tex
\begin{figure}[h]
    \centering
    \includegraphics[width=.5\linewidth]{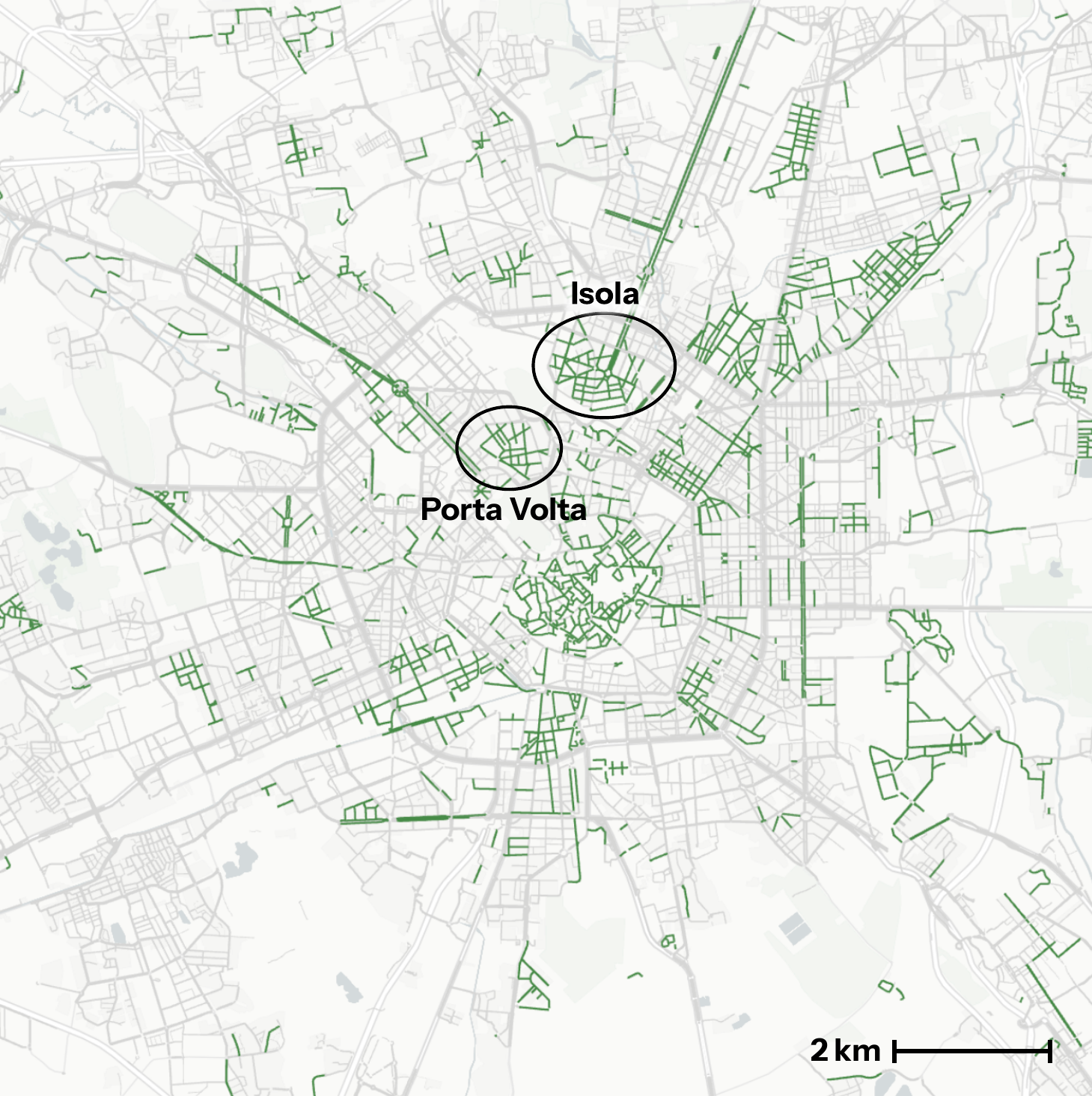}
    \caption{Areas where currently the speed limit is at 30 km/h in Milan city center. Porta Volta is highlighted in red and Isola in blue.}
    \label{fig:case_study_areas} 
\end{figure}

%% file: fig_compliance_map.tex
\begin{figure}[h]
    \centering
    \begin{subfigure}[b]{0.3\linewidth}
        \centering
        \includegraphics[width=0.93\textwidth]{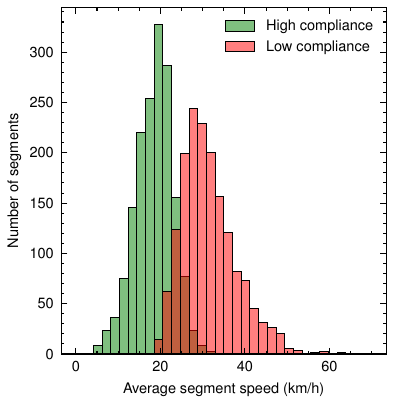}
        \subcaption{\footnotesize Speed distribution of Zones 30 with high and low compliance}
        \label{fig:compliance_speed_comparison}
    \end{subfigure}
    \begin{subfigure}[b]{0.3\linewidth}
            \centering
            \includegraphics[width=0.95\textwidth]{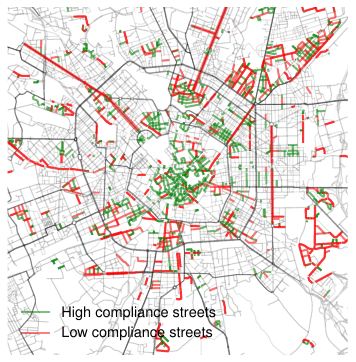}
            \subcaption{\footnotesize Map of the areas with high and low compliance in Zones 30.}
            \label{fig:zones_30_compliance_map}
    \end{subfigure}
    \caption{Speed distribution and compliance map in Zones 30}
    \label{fig:compliance_map}
\end{figure}

%% file: fig_zones_30_features_comparison.tex
\begin{figure*}
    \centering
    \includegraphics[width=\linewidth]{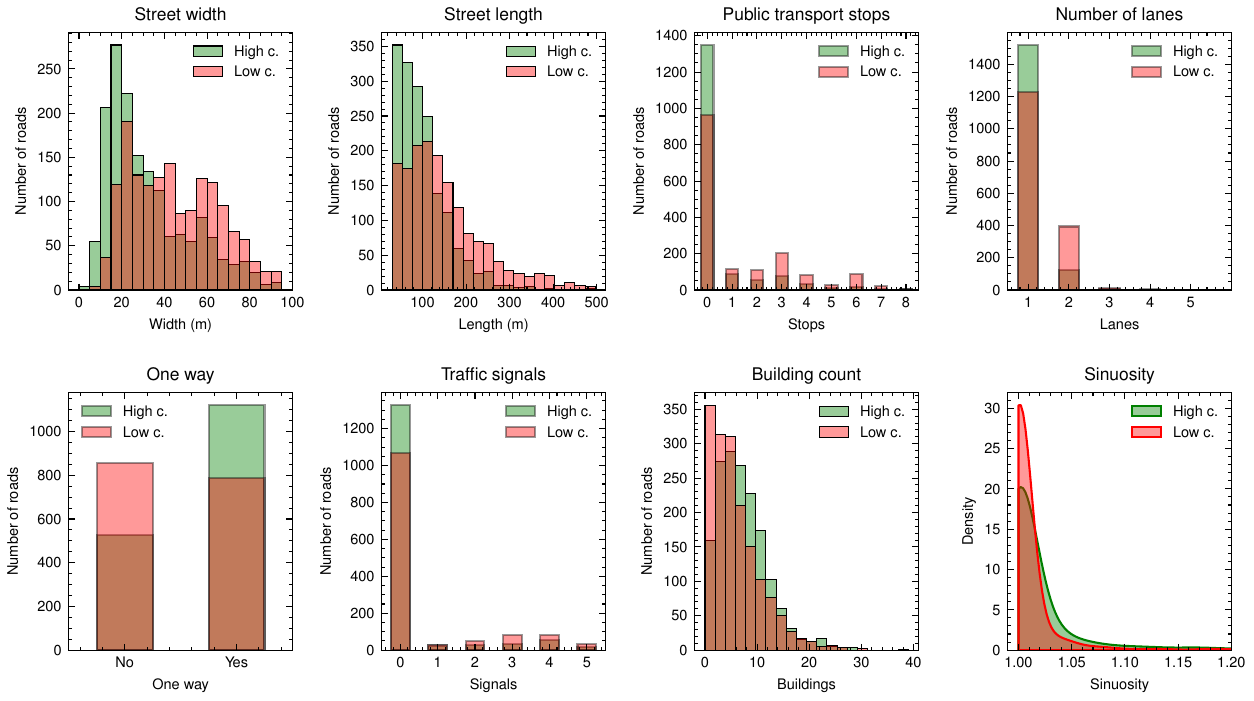}
    \mycaption{OSM features comparison between high and low compliance Zones 30}{We report the features with the lowest $p$-values in the Mann-Whitney U test. All $p$-values are below 0.001. Detailed statistics are reported in \autoref{app:correlation_analysis}.}
    \label{fig:zones_30_osm_features}
\end{figure*}
    
\begin{figure*}
    \centering
    \includegraphics[width=\linewidth]{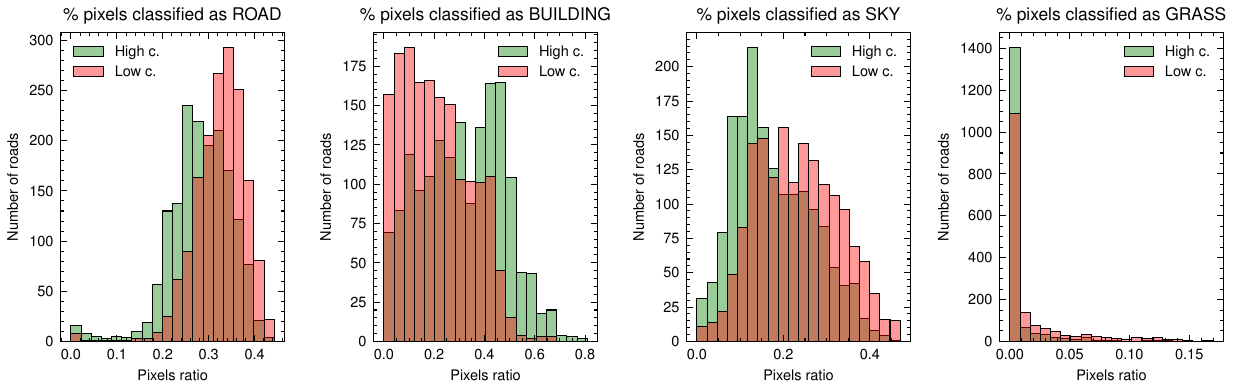}
    \mycaption{GSV features comparison between high and low compliance Zones 30}{We report the features with the lowest $p$-values in the Mann-Whitney U test. All $p$-values are below 0.001. Detailed statistics are reported in \autoref{app:correlation_analysis}.}
    \label{fig:zones_30_gsv_features}
\end{figure*}

%% file: fig_inference_map.tex
\begin{figure}[h]
    \centering
    \includegraphics[width=.5\linewidth]{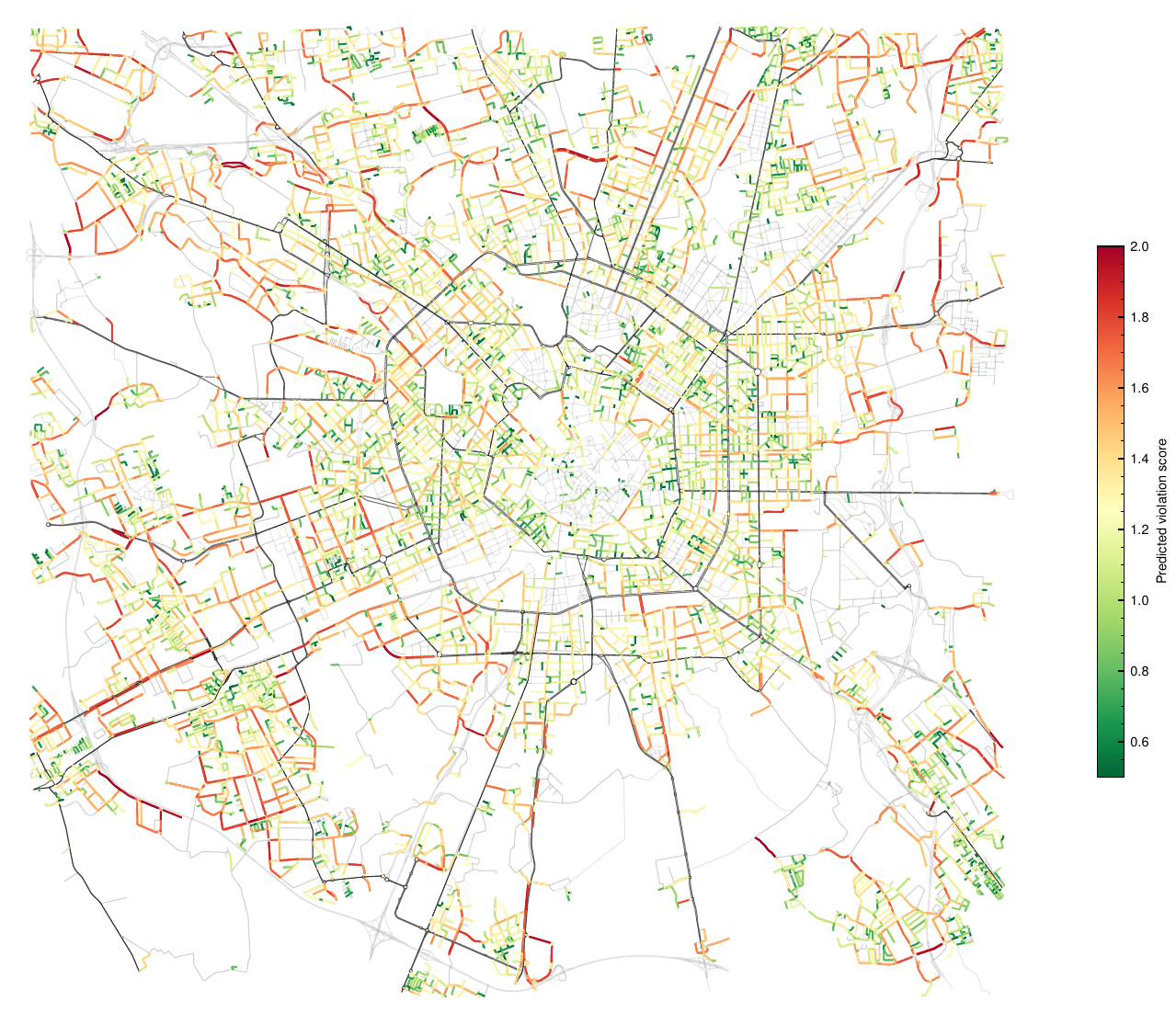}
    \caption{Predicted speed 85th speed percentile with city-wide adoption of limit at 30 km/h.}
    \label{fig:inference_map}
\end{figure}

%% file: fig_inference_images.tex
\begin{figure}
    \centering
    \begin{minipage}{.7\linewidth}\centering
    \begin{subfigure}[b]{0.24\linewidth}
        \centering
        \includegraphics[width=\linewidth]{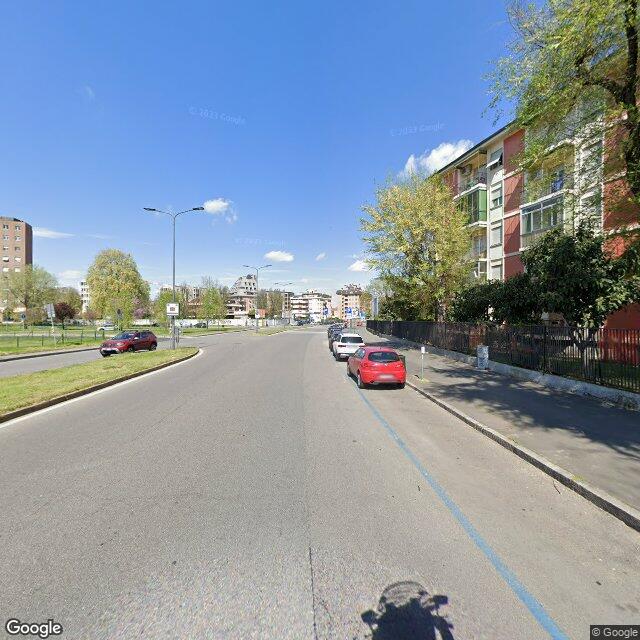}
    \end{subfigure}
    \begin{subfigure}[b]{0.24\linewidth}
        \centering
        \includegraphics[width=\linewidth]{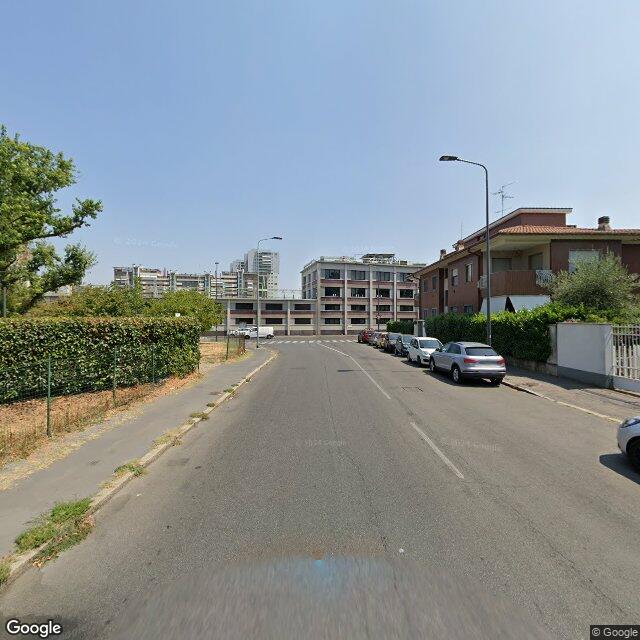}
    \end{subfigure}
    \begin{subfigure}[b]{0.24\linewidth}
        \centering
        \includegraphics[width=\linewidth]{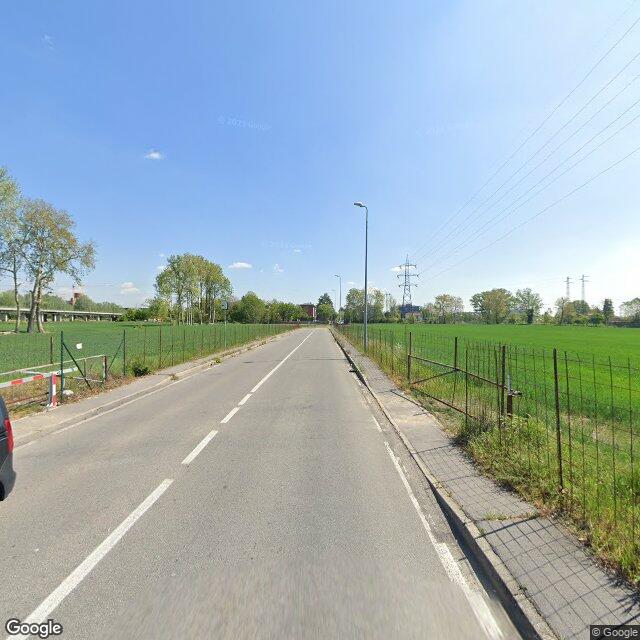}
    \end{subfigure}
    \begin{subfigure}[b]{0.24\linewidth}
        \centering
        \includegraphics[width=\linewidth]{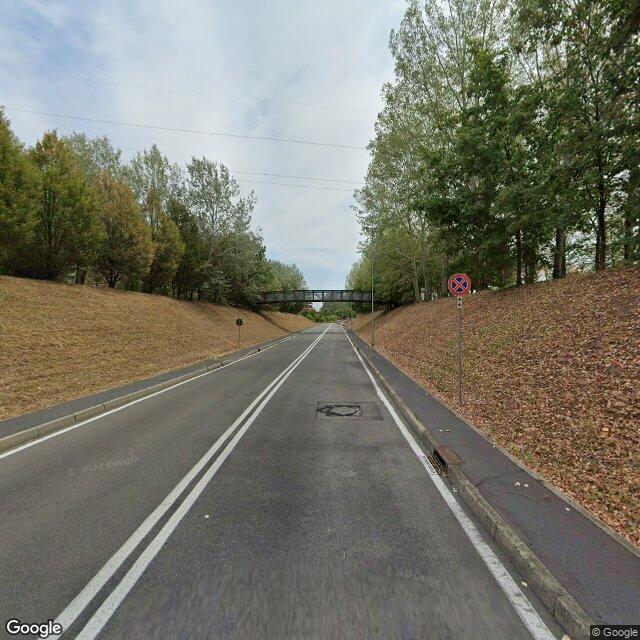}
    \end{subfigure}
    \\[0.1cm]
    \begin{subfigure}[b]{0.24\linewidth}
        \centering
        \includegraphics[width=\linewidth]{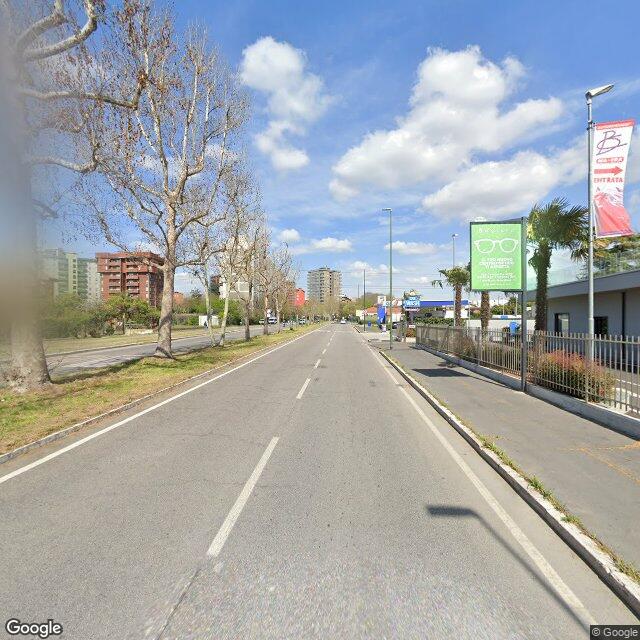}
    \end{subfigure}
    \begin{subfigure}[b]{0.24\linewidth}
    \centering
    \includegraphics[width=\linewidth]{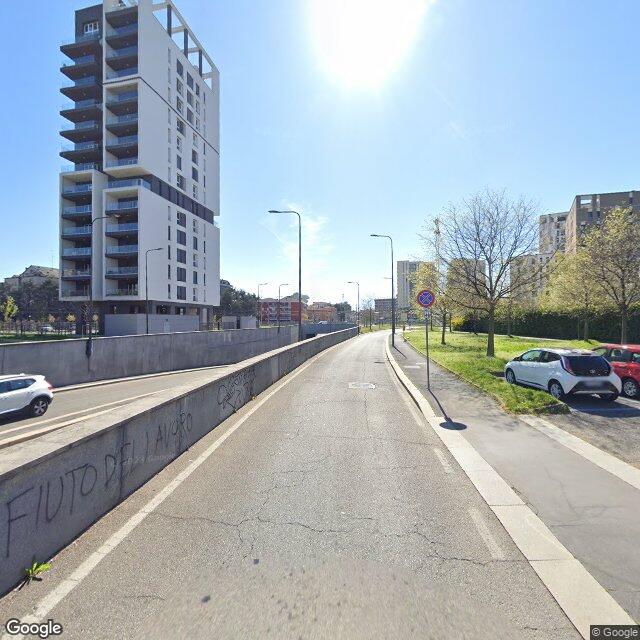}
    \end{subfigure}
    \begin{subfigure}[b]{0.24\linewidth}
    \centering
    \includegraphics[width=\linewidth]{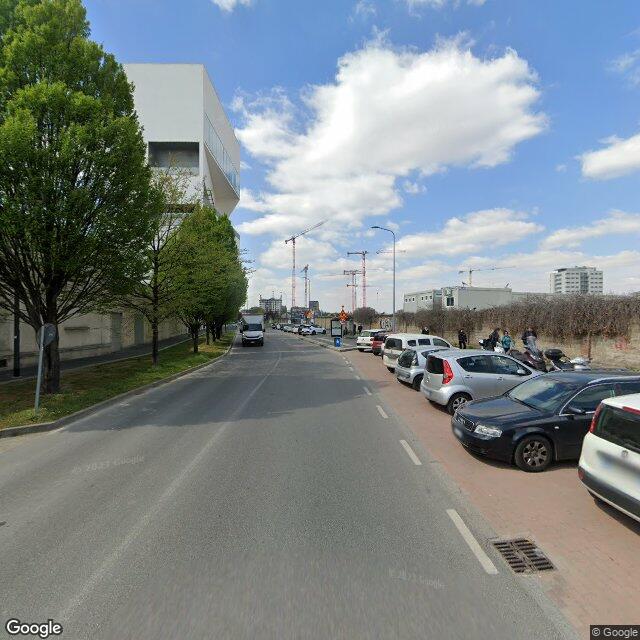}
    \end{subfigure}
    \begin{subfigure}[b]{0.24\linewidth}
        \centering
        \includegraphics[width=\linewidth]{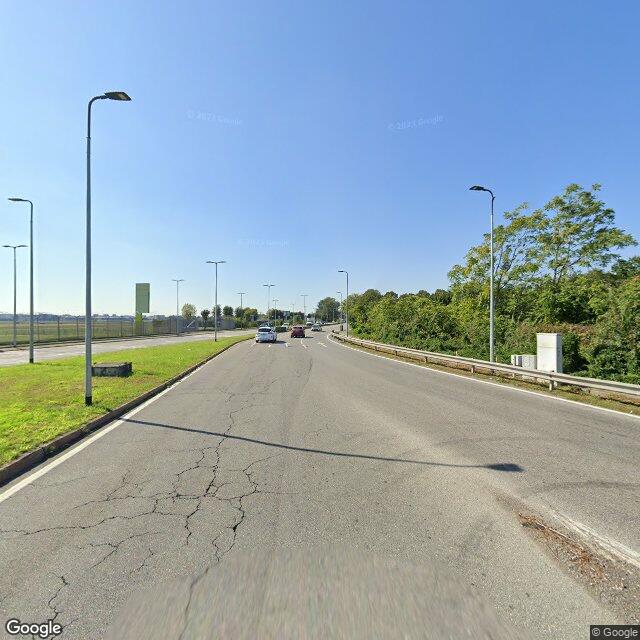}
    \end{subfigure}
    \end{minipage}
    \caption{Randomly sampled streets with predicted low compliance with a 30 km/h limit}
    % \caption{Randomly sampled streets with low predicted 30 km/h compliance}{The images show eight randomly sampled streets with a current limit at 50 km/h that are predicted to have low speed limit compliance if the speed limit was reduced to 30 km/h, as they belong to the top 10\% quantile of the predicted compliance score.}
    \label{fig:inference_high}
\end{figure}

\begin{figure}
    \centering
    \begin{minipage}{.7\linewidth}\centering
    \begin{subfigure}[b]{0.24\linewidth}
        \centering
        \includegraphics[width=\linewidth]{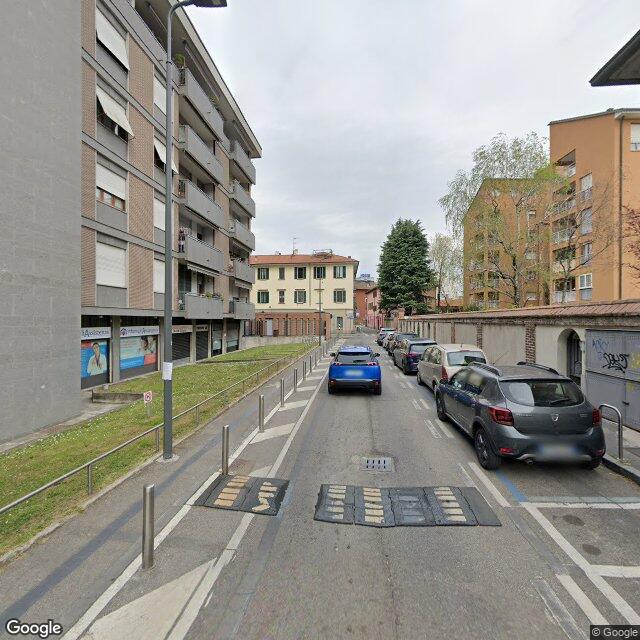}
    \end{subfigure}
    \begin{subfigure}[b]{0.24\linewidth}
        \centering
        \includegraphics[width=\linewidth]{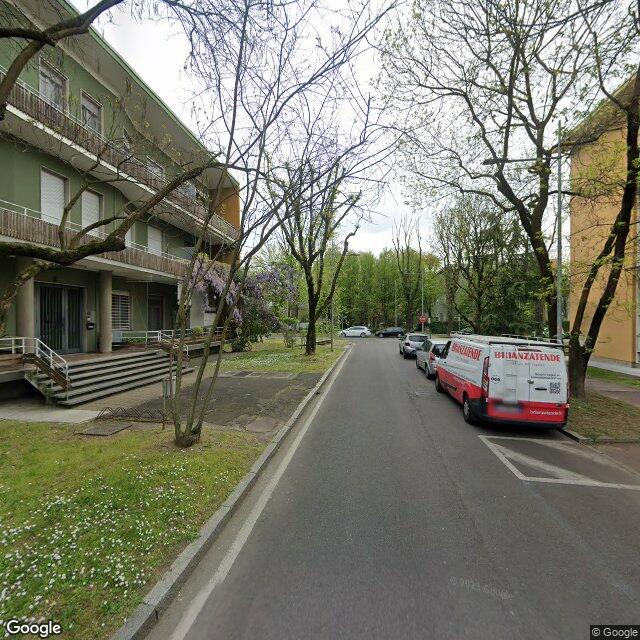}
    \end{subfigure}
    \begin{subfigure}[b]{0.24\linewidth}
        \centering
        \includegraphics[width=\linewidth]{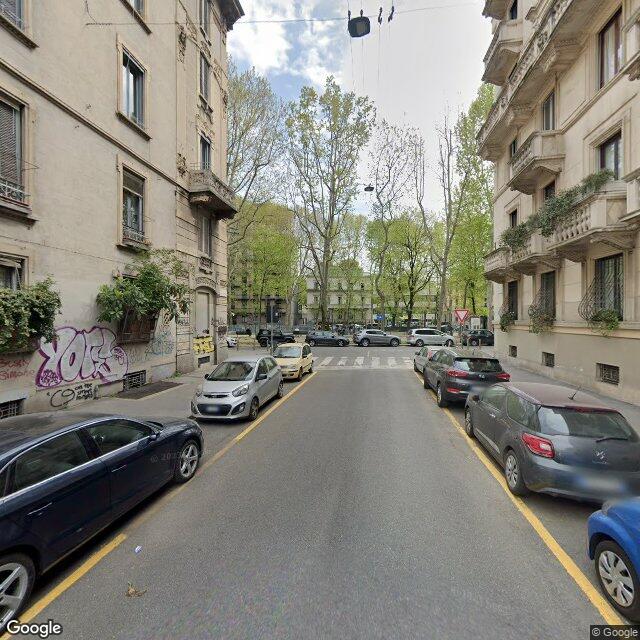}
    \end{subfigure}
    \begin{subfigure}[b]{0.24\linewidth}
        \centering
        \includegraphics[width=\linewidth]{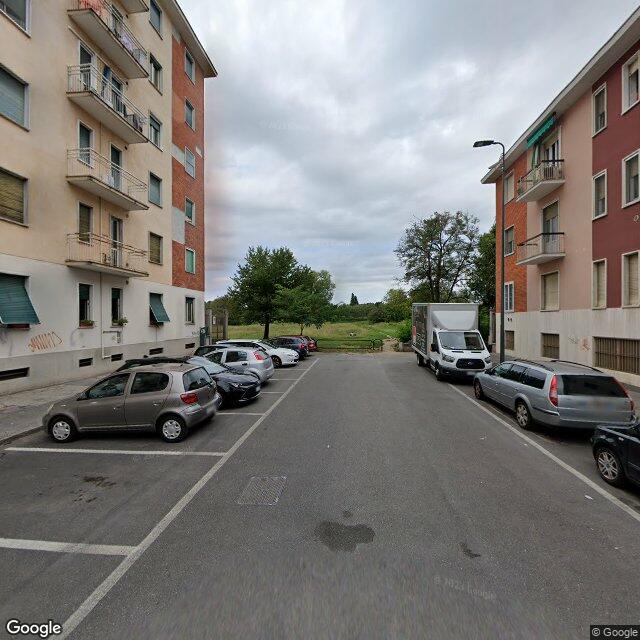}
    \end{subfigure}
    \\[0.1cm]
    \begin{subfigure}[b]{0.24\linewidth}
        \centering
        \includegraphics[width=\linewidth]{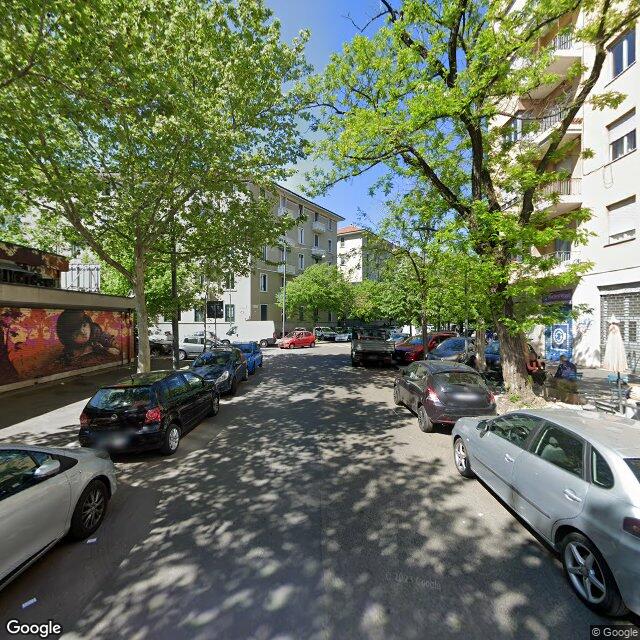}
    \end{subfigure}
    \begin{subfigure}[b]{0.24\linewidth}
    \centering
    \includegraphics[width=\linewidth]{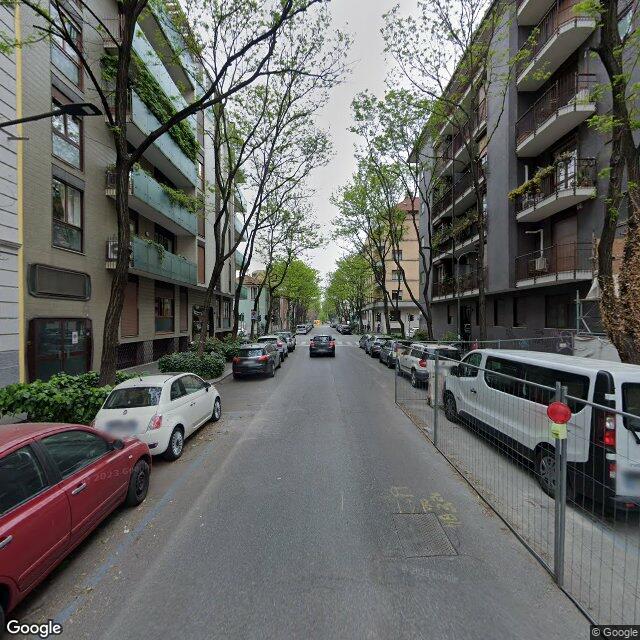}
    \end{subfigure}
    \begin{subfigure}[b]{0.24\linewidth}
    \centering
    \includegraphics[width=\linewidth]{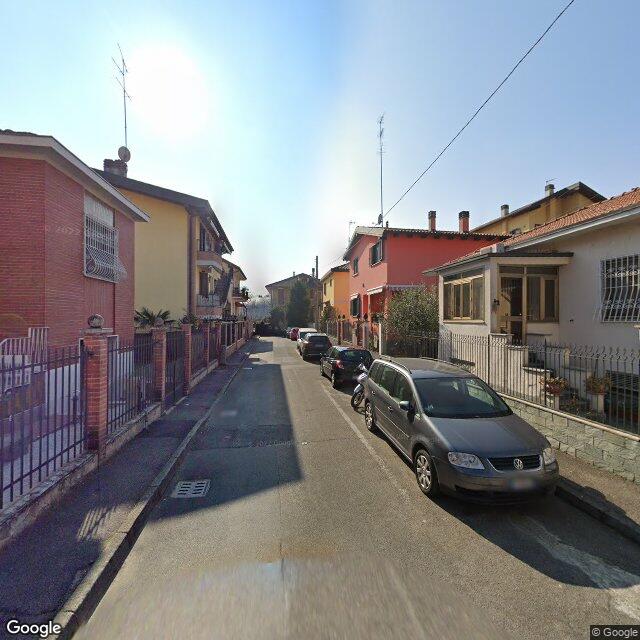}
    \end{subfigure}
    \begin{subfigure}[b]{0.24\linewidth}
        \centering
        \includegraphics[width=\linewidth]{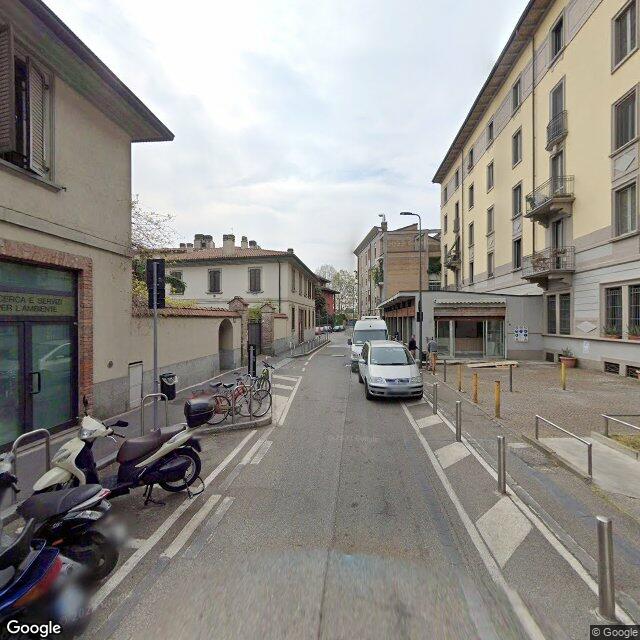}
    \end{subfigure}
    \end{minipage}
    \caption{Randomly sampled streets with predicted high compliance with a 30 km/h limit}
    % \mycaption{Randomly sampled streets with high predicted 30 km/h compliance}{The images show eight randomly sampled streets with a current limit at 50 km/h that are predicted to have high speed limit compliance if the speed limit was reduced to 30 km/h, as they belong to the bottom 10\% quantile of the predicted compliance score.}
    \label{fig:inference_low}
\end{figure}

%% file: discussion.tex
\section{Discussion}
In this study, we first estimated the causal effect of speed limit reductions from 50 km/h to 30 km/h, analyzed the relationship between street characteristics and speed limit compliance, and developed a machine learning model to predict driving speeds based on street features. 

Our causal estimations, through a before-and-after analysis in two areas of Milan where speed limits were reduced, and through a city-wide observational study that matched 30 km/h treated segments with control segments (with limit at 50 km/h) based on similarity of street view images, show that the mere introduction of lower speed limits is not sufficient to reduce driving speeds effectively, highlighting the need to understand how street design – among other strategies – can improve speed limit adherence.

We then investigated the relationship between street characteristics and speed limit compliance, with our findings revealing that narrower streets and denser urban environments contribute to better compliance with limits at 30 km/h, and that multiple lanes and longer road segments are associated with higher speeds, suggesting that lowering speed limits on these roads would require additional interventions to ensure compliance. The effect of one-way streets on compliance is mixed, indicating that policy adjustments should account for various factors. Our findings are consistent with previous studies, which have shown similar results in surveys and simulated environments \cite{qin_quantifying_2020,yao_close_2020,theeuwes_self-explaining_2024}, and extend them to a real-world setting on a city-wide scale. By repeating the analysis in two different local contexts -- Amsterdam and Dubai -- we have shown that the main findings of our study apply to a broad range of cities.

% {\bf is this the first time that these correlations have been highlighted? Or, were already known in qualitative terms -- provide references -- and here quantitatively proven for the first time?}

Finally, we applied machine learning to predict speed limit violations in Milan, which offers a powerful tool for urban planners. 
By forecasting compliance based on street features and other relevant data, urban planning can become more strategic and effective. 
Our model, which achieved an \(R^2\) score of 0.697, shows significant predictive power but also highlights the need for further refinement and the inclusion of additional variables to capture the full complexity of the relationship between street characteristics and drivers' speed. 

Training our model on existing 30 km/h zones and applying it to 50 km/h zones provides valuable insights into where speed limit reductions might be most effective. 
This approach supports more efficient and effective implementation of speed limit policies and helps in planning urban street design interventions. 
The utility of this model extends beyond Milan, as it can be trained in Milan and applied to other cities, which might not have access to the same level of data. 
% However, the transfer learning ability of the model has not been shown. 
By training the model on segmented images rather than raw images, we mitigate the risk of the model overfitting to specific visual characteristics unique to Milan or certain neighborhoods within the city. Deep learning models trained directly on input images might inadvertently learn to associate particular building facades, architectural styles, or other localized features with speed limit compliance, which can severely limit the model's applicability to other areas. 
By extracting generalized features through segmentation, we enhance the model's ability to generalize to different urban settings, potentially increasing its transferability to other cities or regions. This approach reduces the dependency on localized visual cues and emphasizes more universally relevant street characteristics. 
However, the transfer learning ability of the model has not been shown.

% {\bf however, the transfer learning ability of the model has not been shown; we have to mention this as a limitation.} 

The limitations of this study include the potential inaccuracies in the data and the assumptions made in the analysis. OpenStreetMap (OSM) data, while comprehensive, may contain errors or outdated information. 
Although OSM data is generally considered reliable, it is not immune to inaccuracies. 
Our dataset for Milan includes OSM features from a snapshot dated December 31st, 2023, while speed limit compliance data spans from early 2023 to the end of 2023. 
This temporal discrepancy could introduce inaccuracies in the analysis if street features have changed during this period.

Google Street View (GSV) images, obtained from Google's API, always provide the most recent images available: this could mean that some images might have been taken before or after the analyzed period, and that road layout could have changed since then. 
Additionally, our analysis used only a pair of images per street segment, which might not fully capture the street's context. 
Furthermore, the field of view of the captured GSV images is assumed as constant. 
At increasing speeds, the driver's field of view narrows \cite{danielson1956relationship}, and the driver's perception of the road changes. 
We did not account for this in our analysis.

Finally, caution is necessary when interpreting the results as causal relationships. 
The model may capture correlations due to unmeasured confounders, and we did not adjust for potential confounders in comparing street characteristics. 
However, the matching approach used in the causal analysis aimed to minimize the effect of confounders by comparing treated and control segments with similar street features.

% \vspace{3cm}
% \paragraph*{Source code}
% The source code for the methodological pipeline, data processing, and machine learning models is available on GitHub at \url{https://github.com/giacomoorsi/}. Data is not included in the repository it is provided by UnipolTech on license. 

%% file: acknowledgements.tex
\section*{Author Contribution: CRediT}
\textbf{Giacomo Orsi}: Conceptualization, Data curation, Formal analysis, Methodology, Software, Visualization, Writing – original draft, Writing – review and editing. 
\textbf{Titus Venverloo}: Conceptualization, Methodology, Writing – review and editing.
\textbf{Andrea La Grotteria}: Data curation, Formal analysis, Writing – original draft, Writing – review and editing. 
\textbf{Umberto Fugiglando}: Conceptualization, Funding acquisition, Writing – review and editing. 
\textbf{Paolo Santi}: Conceptualization, Supervision, Methodology, Writing – review and editing. 
\textbf{Fábio Duarte}: Conceptualization, Writing – review and editing.
\textbf{Carlo Ratti}: Conceptualization, Funding acquisition, Writing – review and editing.

\section*{Declaration of competing interest}
None.

\section*{Acknowledgments}
The authors would like to thank UnipolTech, AMS Institute, Dubai Future Foundation, and all members of the MIT Senseable City Consortium (Abu Dhabi's Department of Municipalities and Transport, Atlas University, City of Laval, City of Rio de Janeiro, Consiglio per la Ricerca in Agricoltura e l’Analisi dell’Economia Agraria, FAE Technology, Hospital Israelita Albert Einstein, Sondotécnica, Toyota, Volkswagen Group America) for supporting this research. The authors would also like to thank UnipolTech and TomTom for providing the data, and prof. Robert West from EPFL for his advice.

\section*{Data availability}
The authors are not authorized to share the data used in this study.

%% file: appendix.tex
\appendix

\section{Appendix}
\label{chap:appendix}

\subsection{Map Matching}
\label{sec:map_matching}

The speed observations provided by UnipolTech are used to extrapolate a speed profile for each street segment in Milan. A \emph{speed profile} is a set of variables that describes the distribution of speed and the volume of vehicles in a street segment. This approach allows us to reduce the quantity of data points from several millions to a few thousand, as all the hundreds of observations for the same street segments are aggregated into a few variables.

We use the street segments provided by OSM as described in \autoref{sec:osm}. We employ a \emph{map matching} algorithm that associates each speed observation with the corresponding OSM segment. To achieve this, we first sample a point every 3 meters from each segment, obtaining a set of $(x_i, y_i, h_i)$ coordinates, where $(x_i, y_i)$ are the latitude and longitude of the $i$-th point of the segment, projected according to \texttt{EPSG:3003}, and $h_i$ is the angle in degrees (ranging from 0 to 360°) of the point with respect to the following point, as described in \autoref{eq:heading-mapmatching}.

\begin{equation}
\label{eq:heading-mapmatching}
\begin{aligned}
hr_i &= \arctan2(x_{i+1} - x_i, y_{i+1} - y_i) \\
h_i&= 180 \cdot hr_i / \pi
\end{aligned}
\end{equation}

\begin{figure}[h]
    \centering
    \includegraphics[width=0.70\linewidth]{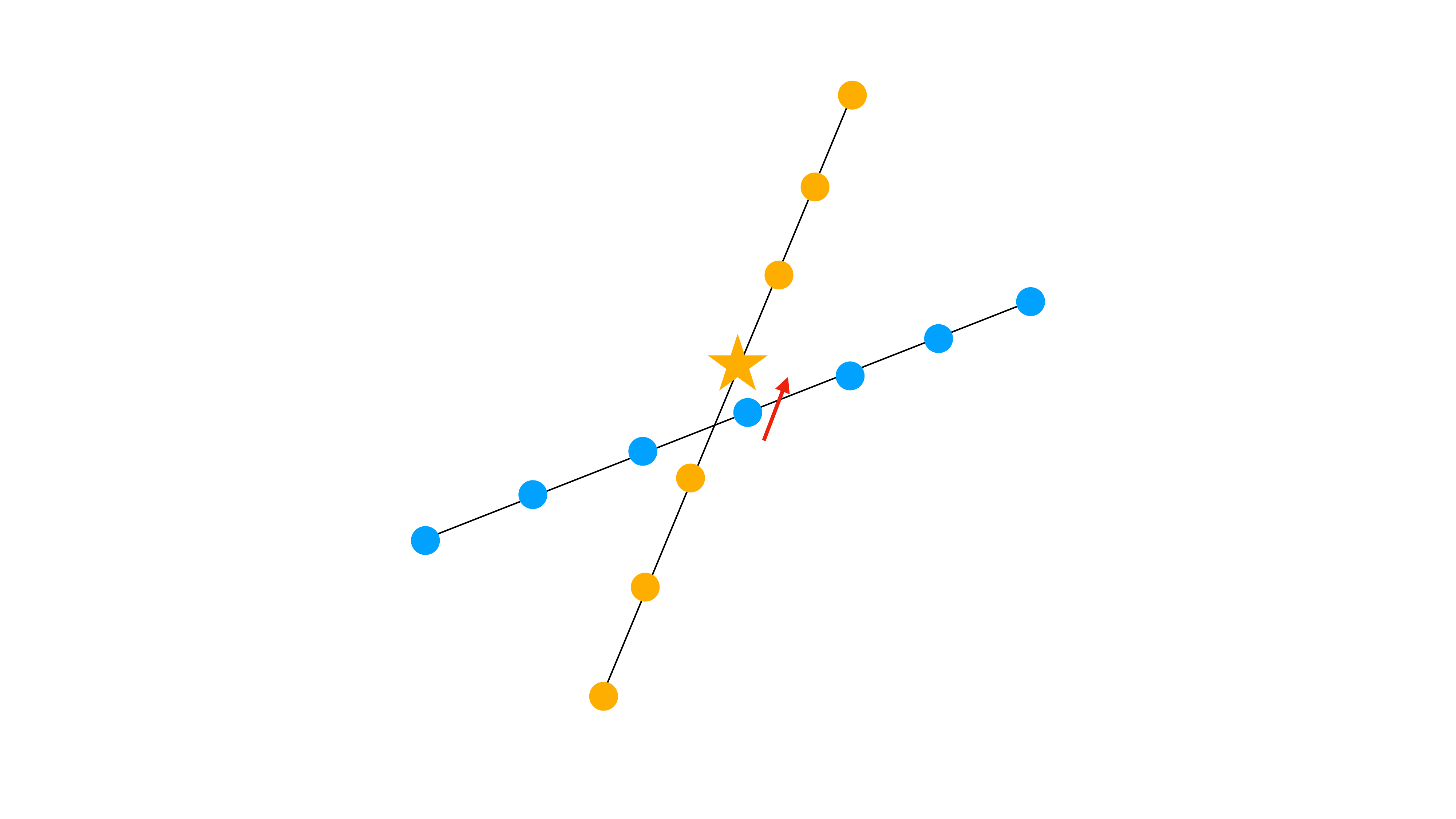}
    \mycaption{Illustration of the map matching algorithm}{The black lines in the figure are two intersecting street segments. We extract a point every 3 meters from the segments. The red arrow is a speed observation with its heading information. The algorithm identifies the 20 points closest to the speed observation, and picks the one with the most compatible heading. The observation is matched with the point visualized with a $\star$. }
    \label{fig:map_matching}
\end{figure}

To associate each speed observation with the corresponding street segment, and given that each observation includes heading information of the vehicles in degrees, we identify the 20 closest $(x_i, y_i, h_i)$ segment points by minimizing the distance of $(x_i, y_i)$ with the car location, and we choose the street segment point with the most compatible heading. Finally, since we are operating in an urban setting with several street segments close to one another, we set two thresholds to ensure we only keep precisely matched observations. Specifically, we discard speed observations with a distance higher than 6 meters from the corresponding matched segment point and those with a heading difference greater than 45 degrees. \autoref{fig:map_matching} shows the benefit of such an algorithm: in that case, a speed observation is not mapped to the closest segment, but rather to a close segment with the most compatible heading.

\subsection{Street View Imagery}
\label{app:google_street_view}

We use Google Street View images to extract visual features of street segments in Milan. The images are obtained using the Google Street View API, which allows the retrieval of images for specific locations and orientations. The images are then processed to extract visual features using computer vision panoptic segmentation techniques. 

The choice to use Google Street View (GSV) images over other community-based services like Mapillary was motivated by the need for standardized, high-quality imagery that ensures consistency across all samples. GSV images are captured using high-quality cameras mounted on Google vehicles, which guarantees a uniform distribution and consistent quality across all images. This level of standardization is crucial when performing image segmentation and comparing pixels classified in each category, as it minimizes variability and enhances the reliability of the extracted features.

Furthermore, GSV provides 360-degree images, allowing for the extraction of images with arbitrary orientations. In our case, this capability allows us to obtain images oriented in both the driving direction and the opposite direction, ensuring comprehensive visual coverage of each street segment. This coherent and standardized approach to image acquisition is fundamental for accurately comparing the visual features of different street segments and drawing robust and interpretable conclusions from the data. 

We extract two images per street segment. Both images are taken from the middle point of each segment, one looking in the direction of the segment and the other in the opposite direction. Formally, to define the heading of each image, we extract the coordinates of the middle point of the segment, $(x_i, y_i)$. Let $(x_i^{+}, y_i^{+})$ and $(x_i^{-}, y_i^{-})$ be the coordinates of the points at 3 meters from the middle point in the direction of the segment and in the opposite direction, respectively. The heading in degrees of the camera for the front image ($h_i^\text{front}$) and the back image ($h_i^\text{back}$) are defined as: 
\[
    h_i^\text{front} = \text{atan2}(x_i^{+} - x_i, y_i^{+} - y_i) \cdot \frac{180}{\pi}
    \]
\[
    h_i^\text{back} = \text{atan2}(x_i^{-} - x_i, y_i^{-} - y_i) \cdot \frac{180}{\pi}
\]

The total number of images to be obtained is $2 \times N$, where $N$ is the number of street segments in the city, which totals 49,346 images. However, the actual number of unique images extracted is 36,460 as some street segments are two-way streets and each direction shares the same images of the other direction. 

The images are extracted using the Google Street View API. The cost of each image is 0.007\$.

We use the settings presented in \autoref{tab:google_street_view_settings} to extract the images. 

\begin{table}[h]
    \centering
    \footnotesize
    \begin{tabular}{ll}
    \toprule
    Setting & Value \\ \midrule
    Pitch & 0 degrees \\
    Field of view & 90 degrees \\
    Image size & 640x640 pixels \\
    Radius & 20 meters \\
    API key & Provided by Google Cloud Platform \\
    API endpoint & \url{maps.googleapis.com/maps/api/streetview} \\
    % API documentation & \url{https://developers.google.com/maps/documentation/streetview} \\
    \bottomrule
    \end{tabular}
    \caption{Settings used to extract Google Street View images.}
    \label{tab:google_street_view_settings}
\end{table}

\subsection{Extraction of OpenStreetMap data}
\label{app:osm_features}
We extract OSM features using a nearby-based approach. As mentioned in \autoref{sec:osm_features}, we compute a buffer of 20 meters around each street segment and extract the features of the OSM elements that intersect with the buffer. OSM features of nearby elements are extracted using the Overpass API. Objects on OpenStreetMap are represented with key-value pairs, where the key is the type of object and the value is the specific attribute of the object. Each object can either be a node (a point or geometry in space) or a way (an ordered list of nodes, which usually represents a line like a street segment)\footnote{See \url{https://wiki.openstreetmap.org/wiki/Map_Features} for a list of OSM key-value pairs.}. Each object can then have a set of key-value pairs called \emph{attributes} that describe the object.
For instance, a building is usually represented as an node with the key \texttt{building} and a value that specifies the type of building (e.g., \texttt{residential}, \texttt{commercial}, etc.) or simply \texttt{yes} if the type is not specified. Some attributes for buildings can include the number of floors, the height, the material, the height, etc.

The Overpass API allows to query OSM data specifying the key-value pairs of interest. The list of the key-value pairs used in this work is shown in \autoref{tab:osm_features_keys}. For some keys, we are interested in counting the number of elements with that key, while for others we are interested in the presence or absence of the key, which is then encoded as a binary feature. For some keys, we are interested in the presence of any value, while for others we only selected a subset of values. 

% split page in two columns and add the two tables
\begin{table}[H]
    \centering
    \footnotesize
    \begin{tabularx}{\linewidth}{llX}
    \toprule
    Type & Key & Value(s) \\ \midrule
    Count & \feature{highway} & \texttt{crossing}, \texttt{traffic\_signals}, \texttt{bus\_stop}, \texttt{pedestrian}, \texttt{cycleway}, \texttt{sidewalk} \\
    Count & \feature{crossing} & \texttt{traffic\_signals}, \texttt{uncontrolled}, \texttt{zebra}, \texttt{marked} \\
    Count & \feature{public\_transport} & \texttt{platform}, \texttt{stop\_position}, \texttt{station} \\
    Count & \feature{amenity} & \texttt{parking}, \texttt{waste\_basket}, \texttt{restaurant}, \texttt{parking\_space}, \texttt{school}, \texttt{fastfood}, \texttt{bycicle\_parking}\\
    Count & \feature{shop} & All values \\
    Count & \feature{building} & All values \\
    Count & \feature{railway} & \texttt{tram\_stop}, \texttt{platform}, \texttt{tram\_crossing}, \texttt{stop}, \texttt{station} \\
    Count & \feature{natural} & \texttt{tree}, \texttt{tree\_row} \\
    Boolean & \feature{highway} & \texttt{footway} \\
    Boolean & \feature{sidewalk} & All values \\
    Boolean & \feature{landuse} & \texttt{grass}, \texttt{residential}, \texttt{industrial}, \texttt{commercial} \\
    Boolean & \feature{footway} & \texttt{sidewalk}, \texttt{crossing} \\
    Boolean & \feature{railway} & \texttt{tram} \\
    Boolean & \feature{cycleway} & \texttt{crossing}, \texttt{lane} \\
    \bottomrule 

    \end{tabularx}
    \caption{OSM features based on the count of elements.}
    \label{tab:osm_features_keys}
\end{table}

We then post-process sidewalk and cycleways features as they require additional processing. Sidewalks can be expressed in OSM as \emph{ways} through the pair \texttt{highway=footway} or \texttt{footway=sidewalk}, but also be represented as an attribute of the street segment itself. In the later case, they are often represented as a \texttt{sidewalk} key with a value that specifies the type of sidewalk (e.g., \texttt{both}, \texttt{left}, \texttt{right}); or they could even be represented with a key \texttt{sidewalk:left} or \texttt{sidewalk:right} and a value as \texttt{yes} or \texttt{no}\footnote{See \url{https://wiki.openstreetmap.org/wiki/Key:sidewalk} for a list of sidewalk key-value pairs.}.
We extract a sidewalk boolean feature if there exists a key which contains \texttt{sidewalk} in the OSM data (regardless if it is a way or an attribute) intersecting the buffer of the street segment, with a value different than \texttt{no}.

Cycleways follow a similar logic. They can be represented as separate object with key \texttt{cycleway} or they can be represented as an attribute of the street segment itself. In the former case, they can be represented with the keys \texttt{cycleway}, \texttt{cycleway:left}, \texttt{cycleway:right}, \texttt{cycleway:both}, and many others\footnote{See \url{https://wiki.openstreetmap.org/wiki/Key:cycleway} for a list of cycleway key-value pairs.}. We extract a cycleway boolean feature if there exists a key which contains \texttt{cycleway} in the OSM data (regardless if it's a way or an attribute) intersecting the buffer of the street segment, with a value different than \texttt{no}.

% paragraph about removing: 
% "amenity=parking_space", "bridge", "highway=emergency_bay", "highway=living_street"
% tunnel=*
% junction=*
From the obtained OSM features, we remove those that were present in less than 5\% of the street segments, as they are not informative for the analysis. Those are the following key-value pairs: \texttt{amenity=parking\_space}, \texttt{bridge=*}, \texttt{highway=emergency\_bay}, \texttt{highway=living\_street}, \texttt{tunnel=*}, and \texttt{junction=*} (where * is a wildcard that represents any value).

Finally, we merge together many keys that are related to public transport facilities into two features: \texttt{public\_transport\_stop} and \texttt{railway\_tram}.

\subsection{Results Appendix}
\label{app:results}

\subsubsection{Correlation analysis of features in Zones 30}
\label{app:correlation_analysis}

In \autoref{tab:most_significant_features} we report the 30 features with the lowest $p$-values given by the Mann-Whitney U test in the comparison of high and low compliance zones 30, as described in \autopageref{sec:correlation_analysis}. The table also includes Cohen's $d$ effect size, defined as the difference between the means of the two groups ($d$) divided by the pooled standard deviation. The last two columns show the mean value of the feature in the high and low compliance zones, respectively ($\mu_{h}$ and $\mu_{l}$).

% A map with the location of the areas with high and low compliance is shown in \autoref{fig:zones_30_compliance_map}.

\begin{table*}
    \centering
    \footnotesize
    % \begin{tabular}{|l|c|c|c|c|c|c|}
    \begin{tabular}{lcccccc}
    \toprule
    Feature & Category & $p$-value & Cohen's $d$ & $d$ & $\mu_{h}$ & $\mu_{l}$ \\ \midrule
    \feature{building\_width} & OSM & 6.372e-73 & -0.564 & -12.901 & 33.855 & 46.756 \\ 
    \feature{pixels\_construction--structure--building} & Vision & 2.220e-69 & 0.628 & 0.104 & 0.293 & 0.188 \\ 
    \feature{length} & OSM & 5.235e-66 & -0.614 & -49.168 & 99.982 & 149.150 \\ 
    \feature{highway\_bus\_stop} & Nearby street & 7.997e-58 & -0.513 & -0.305 & 0.092 & 0.397 \\ 
    \feature{public\_transport\_stop} & OSM & 5.534e-52 & -0.492 & -0.924 & 0.520 & 1.444 \\ 
    \feature{pixels\_construction--flat--road} & Vision & 5.938e-50 & -0.498 & -0.032 & 0.223 & 0.255 \\ 
    \feature{pixels\_nature--terrain} & Vision & 1.023e-45 & -0.397 & -0.010 & 0.007 & 0.017 \\ 
    \feature{lanes} & OSM & 2.141e-42 & -0.460 & -0.193 & 1.080 & 1.273 \\ 
    \feature{pixels\_nature--sky} & Vision & 1.428e-40 & -0.464 & -0.044 & 0.164 & 0.208 \\ 
    \feature{highway=residential} & Street type & 1.507e-40 & 0.465 & 0.223 & 0.748 & 0.525 \\ 
    \feature{oneway} & OSM & 4.740e-32 & 0.411 & 0.203 & 0.681 & 0.479 \\ 
    \feature{highway=tertiary} & Street type & 1.729e-30 & -0.400 & -0.115 & 0.033 & 0.148 \\ 
    \feature{pixels\_marking--general} & Vision & 5.267e-28 & -0.331 & -0.005 & 0.013 & 0.018 \\ 
    \feature{highway\_traffic\_signals} & Nearby street & 5.271e-26 & -0.335 & -0.329 & 0.276 & 0.605 \\ 
    \feature{crossing\_traffic\_signals} & OSM & 3.932e-25 & -0.336 & -1.075 & 1.078 & 2.153 \\ 
    \feature{building\_count} & OSM & 3.157e-23 & 0.275 & 1.385 & 7.007 & 5.621 \\ 
    \feature{pixels\_nature--vegetation} & Vision & 3.387e-22 & -0.298 & -0.034 & 0.095 & 0.129 \\ 
    \feature{pixels\_object--vehicle--motorcycle} & Vision & 1.793e-18 & 0.170 & 0.001 & 0.003 & 0.002 \\ 
    \feature{pixels\_construction--barrier--guard-rail} & Vision & 4.764e-18 & -0.135 & -0.001 & 0.000 & 0.001 \\ 
    \feature{sinuosity} & OSM & 1.064e-14 & 0.162 & 0.017 & 1.036 & 1.019 \\ 
    \feature{pixels\_construction--structure--bridge} & Vision & 3.140e-14 & -0.071 & -0.002 & 0.001 & 0.003 \\ 
    \feature{pixels\_human--person} & Vision & 4.106e-14 & 0.234 & 0.001 & 0.002 & 0.001 \\ 
    \feature{crossing\_uncontrolled} & OSM & 1.668e-13 & 0.269 & 0.880 & 4.311 & 3.431 \\ 
    \feature{pixels\_object--support--utility-pole} & Vision & 5.726e-13 & -0.154 & -0.000 & 0.000 & 0.001 \\ 
    \feature{cycleway} & OSM & 1.182e-12 & -0.248 & -0.110 & 0.213 & 0.322 \\ 
    \feature{amenity\_waste\_basket} & OSM & 2.764e-12 & 0.183 & 0.210 & 0.615 & 0.404 \\ 
    \feature{pixels\_construction--flat--sidewalk} & Vision & 3.433e-12 & 0.250 & 0.009 & 0.038 & 0.029 \\ 
    \feature{pixels\_construction--flat--rail-track} & Vision & 3.506e-12 & -0.182 & -0.005 & 0.006 & 0.011 \\ 
    \feature{pixels\_object--vehicle--car} & Vision & 3.568e-12 & 0.273 & 0.014 & 0.075 & 0.060 \\ 
    \feature{highway\_pedestrian} & Nearby street & 4.722e-12 & 0.234 & 0.223 & 0.455 & 0.232 \\ 
    \bottomrule
    \end{tabular}
    \caption{Features with most significant change in the comparison of high and low compliance Zones 30.}
    \label{tab:most_significant_features}
\end{table*}

% \begin{figure}[b]
%     \centering
%     \includegraphics[width=0.7\textwidth]{results/data_analysis/zones_30_map_compliance.pdf}
%     \caption{Map of the areas with high and low compliance in Zones 30.}
%     \label{fig:zones_30_compliance_map}
% \end{figure}

\subsubsection{Machine learning models}
\label{app:machine_learning_models}

We fit machine learning models to predict the compliance score over 22,876 street segments, as detailed in \autoref{sec:machine_learning}. We use linear regression, random forest, gradient boosting models, and neural networks. We report the hyper-parameters used for each model and performance metrics in \autoref{tab:ml_models_hyperparameters} and \autoref{tab:ml_models_hyperparameters_withoutnobs}. The hyper-parameters are selected using a grid search with 5-fold cross-validation. The performance metrics are the mean absolute error (MAE), the mean squared error (MSE), and the $R^2$ score. The best model is the gradient boosting model, which achieves an $R^2$ score of 0.697. 

% We remove the segmentation classes with an average low number of classified pixels and we only keep the following most relevant classes: road, sky, building, tree, car, sidewalk, grass \& fence. 

\begin{table*}
    \centering
    \tablesize  
    \begin{tabularx}{\linewidth}{llXccc}
    \toprule
    Model & Training data & Hyper-parameters & MAE & $R^2$ train & $R^2$ test \\ \midrule
    Least squares & all data & - & 0.160  & 0.543 & 0.542 \\
    Least squares & zones 30 & - & 0.217 & 0.485 & 0.468 \\
    Neural network & all data & \texttt{activation=logistic, alpha=0.0001, hidden\_layer\_sizes=(100, 100), learning\_rate=0.005} & 0.137  & 0.706 & 0.664 \\
    Neural network & zones 30 & \texttt{activation=logistic, alpha=0.0001, hidden\_layer\_sizes=(100, 100, 100), learning\_rate=0.01} & 0.189 & 0.647 & 0.540 \\
    Gradient Boosting & all data & \texttt{loss=absolute\_error, max\_depth=5, min\_samples\_leaf=1, min\_samples\_split=10, n\_estimators=1000} & 0.119 & 0.802 & \textbf{0.719} \\

    Gradient Boosting & zones 30 & \texttt{loss=absolute\_error, max\_depth=7, min\_samples\_leaf=4, min\_samples\_split=10, n\_estimators=500} & 0.174 & 0.919 & \textbf{0.646} \\    

    \bottomrule
    \end{tabularx}
    \caption{Predictive models, including the approximate traffic density feature.}
    \label{tab:ml_models_hyperparameters}
\end{table*}

\begin{table*}
    \centering
    \tablesize  
    \begin{tabularx}{\linewidth}{llXccc}
    \toprule
    Model & Training data & Hyper-parameters & MAE & $R^2$ train & $R^2$ test \\ \midrule
    Gradient Boosting & all data & \texttt{loss=absolute\_error, max\_depth=5, min\_samples\_leaf=1, min\_samples\_split=10, n\_estimators=1000} & 0.139 & 0.737 & \textbf{0.621} \\
    Gradient Boosting & zones 30 & \texttt{loss=absolute\_error, max\_depth=7, min\_samples\_leaf=4, min\_samples\_split=10, n\_estimators=500} & 0.202 & 0.870 & \textbf{0.534} \\    
    \bottomrule
    \end{tabularx}
    \caption{Predictive models, excluding the approximate traffic density feature.}
    \label{tab:ml_models_hyperparameters_withoutnobs}
\end{table*}